%% file: TSG_R0.tex
\documentclass[journal]{IEEEtran}
\usepackage{color,amsmath,amstext,amsfonts,amssymb, mathrsfs,mathtools}
\usepackage{graphicx,algorithm,algorithmic}
\usepackage{amsthm}

\usepackage{comment}
\usepackage{graphicx}
 \newtheorem{theorem}{Theorem}
 
 \newtheorem{lemma}{Lemma}

 \newtheorem{definition}[theorem]{Definition}
 
\usepackage{amssymb}

\usepackage{multirow}
\usepackage{cite}
\usepackage{graphics}
\ifCLASSINFOpdf
\else
\fi

\usepackage{url}
\usepackage{mathrsfs}
\usepackage{comment}

\usepackage[dvipsnames,svgnames]{xcolor}
\usepackage{bbm}

\begin{document}

\title{Scalable Grid-Aware Dynamic Matching using Deep Reinforcement Learning}
\author{Majid Majidi \IEEEmembership{Student Member, IEEE}, Deepan Muthirayan \IEEEmembership{Member, IEEE}, Masood Parvania \IEEEmembership{Senior Member, IEEE}, Pramod P. Khargonekar \IEEEmembership{Fellow, IEEE}
\vspace{-10pt}
\thanks{
This work is supported in part by the National Science Foundation under Grant ECCS-1839429. 

M. Majidi, and M. Parvania are with the Department of Electrical and Computer Engineering, the University of Utah, Salt Lake City, UT 84112 USA (e-mails: \{majid.majidi, masood.parvania,\}@utah.edu). Deepan Muthirayan and Pramod P. Khargonekar are with the Department of Electrical Engineering and Computer Science, University of California, Irvine, CA 92697 USA (e-mails: \{deepan.m, pramod.khargonekar,\}@uci.edu).}}

\input{macros}

\maketitle
\begin{abstract}
This paper proposes a two-level hierarchical matching framework for Integrated Hybrid Resources (IHRs) with grid constraints. An IHR is a collection of Renewable Energy Sources (RES) and flexible customers within a certain power system zone, endowed with an agent to match. 
The key idea is to pick the IHR zones so that the power loss effects within the IHRs can be neglected. This simplifies the overall matching problem into independent IHR-level matching problems, and an upper-level optimal power flow problem to meet the IHR-level upstream flow requirements while respecting the grid constraints. 
Within each IHR, the agent employs a scalable Deep Reinforcement Learning algorithm to identify matching solutions such that the customer's service constraints are met. 
The central agent then solves an optimal power flow problem with the IHRs as the nodes, with their active power flow and reactive power {capacities}, and grid constraints to scalably determine the final flows such that matched power can be delivered to the extent the grid constraints are satisfied. 
The proposed framework is implemented on a test power distribution system, and multiple case studies are presented to substantiate the welfare efficiency of the proposed solution and the satisfaction of the grid and customers' servicing constraints.     
\end{abstract}


\IEEEpeerreviewmaketitle
\section{Introduction}
\IEEEPARstart{D}{riven} by the advances in communication technologies and supporting policies, Distributed Energy Resources (DERs) and flexible loads are going to be highly penetrated in power grids.
Federal Energy Regulatory Commission (FERC) order 2222 requires power system operators to facilitate the participation of demand-side resources in the electricity markets, reflecting their significant potential to provide energy flexibility \cite{no2222}.
The DERs and flexible loads, if coordinated and controlled carefully, can make the grid flexible and energy-efficient \cite{oikonomou2019deliverable, oikonomou2020coordinated}.
However, a large number of DERs and flexible loads might challenge the structure and capacity of power distribution grids.
Hence, it is essential to develop intelligent energy management solutions that can manage different sorts of DERs and flexible loads in a scalable manner without compromising the power grid's stability.

In recent years, several efforts have been made to develop energy management solutions for the coordination of DERs in power distribution systems. 
One promising solution is matching, which is a Peer-to-Peer (P2P) solution.
Unlike traditional energy management solutions based on pooling resources, matching offers optimal use of energy flexibility while accounting for the energy preferences of flexible customers.
This is the key feature that makes matching very promising to future power grids.
However, developing a matching solution for power grids still has several challenges: (i) the solution has to be online and capable of adapting the matching strategy with the state of the whole grid, (ii) it has to cope with a large number of DERs and flexible loads, and (iii) it has to satisfy the security constraints of the grid.
Although \cite{muthirayan2020online, majidi2021dynamic} propose online solutions for dynamic matching, the proposed solutions are heuristics and, therefore, can be sub-optimal. 
Alternatively, data-driven approaches like Reinforcement Learning (RL) can be used to discover high-performing dynamic matching policies.
However, RL approaches have severe limitations when it comes to learning policies for power grids with constraints such as power flow limits.
For instance, \cite{majidi2021dynamic1} proposes a Deep Reinforcement Learning (DRL) solution for dynamic matching, but it fails to account for the grid constraints.
Moreover, a central matching structure might not be efficient or feasible for managing a large number of DERs and flexible loads in the power grid.
Hence, an efficient learning-based hierarchical matching model with Integrated Hybrid Resources (IHRs) is of interest in addressing dynamic matching in power grids with grid constraints.

IHRs present a viable solution to facilitate efficient energy management and control of uncertain DERs in various applications \cite{IHR_Report1, IHR_Report, majidi2022risk, li2022hybrid}. 
In a power grid, different types of renewable and non-renewable DERs and flexible loads can be combined and operated as an IHR to supply distributed energy flexibility.
The key feature of an IHR is that  
the resources within the IHR can be treated as an integrated set of resources with a single interconnection point. 
Therefore, from a matching solution point of view, each IHR can be treated as a separate matching market with a single interconnection to the distribution grid. This then enables the use of an RL-type algorithm to determine the optimal way to manage the DERs within each IHR, where the RL solution does not need to take into account the grid constraints.
Once the IHR-level matching results are determined, a central controller can {adjust the IHR setpoints} using a reduced-dimension Optimal Power Flow (OPF) model with each IHR as a node to balance the excesses while ensuring that the grid constraints across the distribution system are satisfied.
{The adjusted setpoints are then communicated to re-dispatch the DERs and flexible loads in each IHR.}

\begin{figure}[ht]
\centering{
\includegraphics[width=1\linewidth]{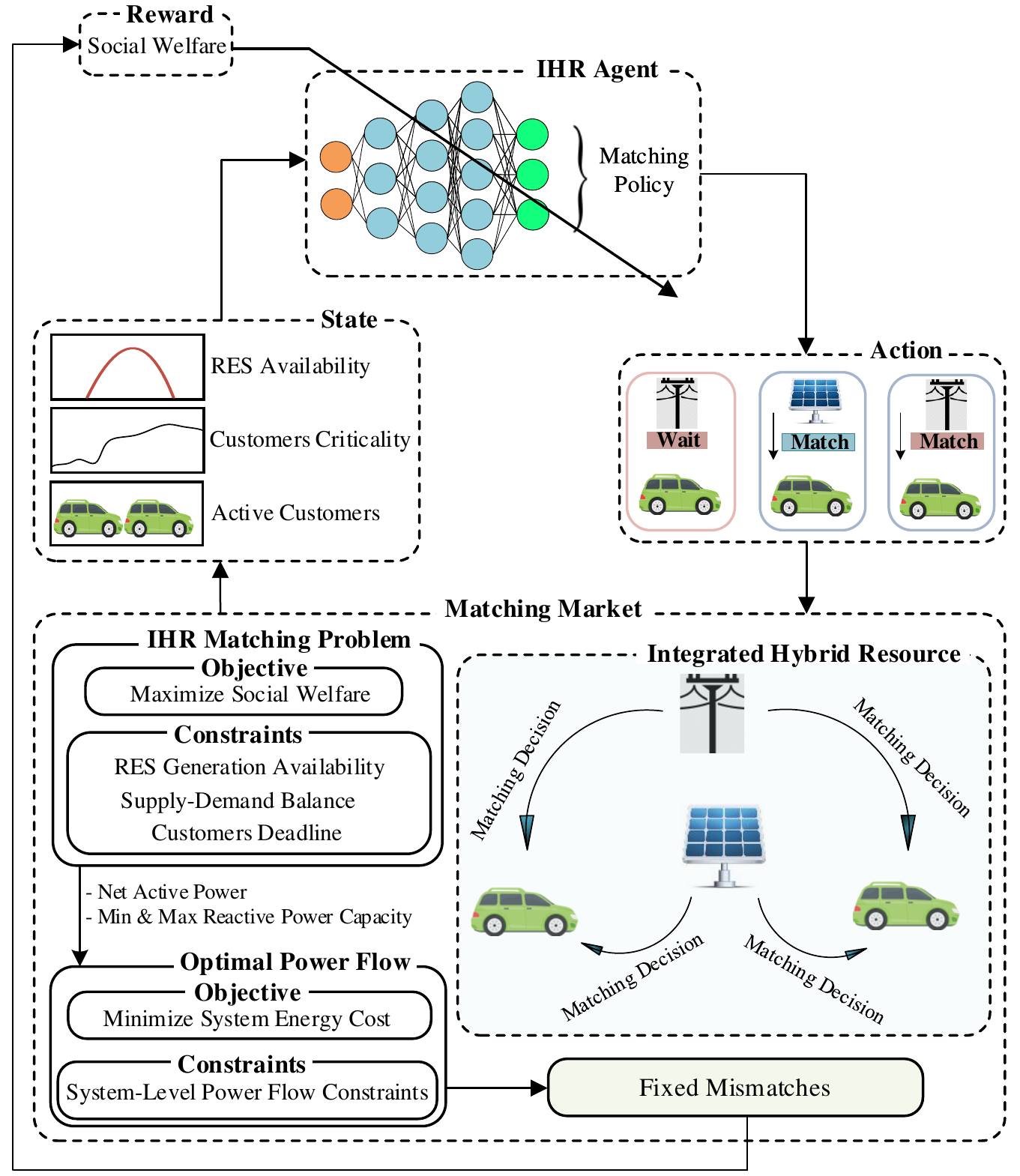}}
\caption{Structure of hierarchical dynamic matching model in power distribution systems using deep reinforcement learning.}
\label{fig:Structure}
\end{figure}

This paper proposes a hierarchical framework for dynamic matching markets in power distribution systems composed of IHRs. 
The schematic of the proposed framework is shown in Fig. \ref{fig:Structure}.
An IHR consists of an agent that employs DRL to locally match the distributed Renewable Energy Sources (RES) and flexible customers in each IHR while satisfying the quality of service constraints of customers, i.e., criticality, servicing deadline, etc. 
Such a learning-based approach allows for developing a very effective online matching policy, which is otherwise very difficult to design. 
Once the IHR-level matching results are determined, each IHR agent communicates the net active power flow (i.e., net active power consumption/generation), as well as the reactive power {capacities} of the IHR to a central agent.
The central agent then formulates a reduced-dimension OPF model with the IHRs as the nodes to determine the final flows (setpoints) such that the grid constraints are met.
In this stage, the central agent can curtail the IHR-level matching decisions and control the reactive power flow from/into each IHR in order to make sure that power flows in the grid and voltage levels across the distribution nodes are not violated.

\subsection{Related Works and Contributions}
Several works in the literature have explored the control and management of DERs in distribution systems.
P2P energy trading markets of different types are developed in \cite{tushar2021peer, luth2018local, morstyn2018bilateral, guerrero2018decentralized, liu2022fully, botelho2022integrated, soriano2021peer, amin2020motivational, luo2022distributed, zheng2022peer}.
The authors in \cite{luth2018local} studied the operation and benefits of centralized and decentralized battery energy storage under different P2P market designs.
A P2P market design based on bilateral contracts is proposed in \cite{morstyn2018bilateral} for energy trading between multiple DERs and flexible loads, with the objective of minimizing peak load in the low-voltage power distribution system. 
A decentralized P2P optimization framework is presented in \cite{guerrero2018decentralized} to enable local energy sharing between DER agents in the low-voltage distribution systems. 
A local energy sharing framework is proposed for prosumers in low-voltage distribution systems in \cite{liu2022fully}, where the voltage regulation capability of the proposed energy sharing framework is highlighted. 
An iterative sequential approach is implemented in \cite{botelho2022integrated} to enable P2P energy and reserve sharing between prosumers in power distribution systems with grid constraints. 
A negotiation algorithm is proposed to facilitate energy sharing between interconnected DER owners in \cite{soriano2021peer}.  
The use of game theory to determine the interaction strategy of DERs in P2P energy sharing is investigated in \cite{ amin2020motivational, luo2022distributed, zheng2022peer}.

The application of data-driven approaches to energy management of DERs is studied in \cite{fang2021multi, li2021peer, wang2021surrogate, ye2021scalable, qiu2021scalable}. 
The authors in \cite{fang2021multi} implemented a multi-agent learning framework to determine real-time local energy trading strategies for DER owners in regional microgrids. 
A multi-energy sharing model based on RL is proposed in \cite{li2021peer, wang2021surrogate} for local heat and power sharing in energy microgrids equipped with DERs. 
In \cite{ye2021scalable, qiu2021scalable}, the authors proposed a specific price-based market framework for coordinating the prosumers in the market to minimize the peak load. 
In \cite{hosseini2022hierarchical, hosseini2023hierarchical}, hierarchical energy management models based on DRL are proposed for local energy management of energy storage systems to improve the resilience of the power distribution system.

Although the works reviewed here study the management and coordination of DERs in distribution systems, their P2P solutions address specific scenarios, and many do not account for the grid constraints. 
Moreover, the existing hierarchical energy management models for DERs in power distribution systems neglect the preference and dynamic characteristics of the DERs and flexible loads, which impacts the energy flexibility available to the grid. 
In contrast, this paper develops a broadly applicable online matching solution that (i) is designed to maximize the integration of local RES and the overall welfare in a very generic real-time operating scenario that can be riddled with uncertainties, (ii) takes into consideration the flexible loads' servicing deadline, as well as their dynamic criticality, and willingness to pay for a unit of energy, and (iii) at the same time accounts for the power grid constraints. 
The key contributions of the paper can be summarized as follows:
\begin{itemize}
    \item A hierarchical dynamic matching framework for power distribution systems constituted by IHRs. The hierarchical framework allows for deploying the combined leaning-based and optimization solutions and their advantages for scalable dynamic matching in power grids.
    \item An efficient and scalable learning-based solution to match the flexible loads and uncertain RES within each IHR with no need for prior experience or expert supervision, or elaborate design.
    An independent DRL algorithm for each IHR agent that matches the flexible loads and supply sources to improve the utilization of RES and, therefore, maximize the social welfare in each IHR such that the quality of service constraints of the loads are satisfied prior to their servicing deadline, and their dynamic willingness to pay for a unit of energy is taken into account. 
    \item An upper-level optimization model to fix the excesses or imbalances within the IHRs.
    Once the matching results in each IHR are determined, a central agent runs a reduced-dimension OPF problem to adjust the setpoints in a scalable manner and fix the possible imbalances within the IHRs, and, at the same time, ensure the power flow limits and voltage security constraints of the distribution system are met. Individual rewards (i.e., social welfare) along with adjusted setpoints are then communicated with the IHRs at each time to re-dispatch the local resources.
    Numerical studies verify that running the upper-level model is essential to minimize the curtailment of local matching decisions due to the violation of the grid's constraints over time. 
    
\end{itemize} 

The rest of the paper is categorized as follows: Hierarchical matching market framework is presented in Section \ref{Matching Market}. The proposed learning-based solution approach is explained in Section \ref{DRL}. The simulation results are presented in Section \ref{Simulations}, and the paper is concluded in Section \ref{Conclusions}.

\allowdisplaybreaks
\section{Hierarchical Matching Market} \label{Matching Market}
The proposed hierarchical matching framework is composed of (i) a {\it market operator or central agent}, and (ii) multiple IHRs, with each IHR operating as a separate matching market and the central agent acting as the coordinating agent between the IHRs. 
In the proposed framework, the grid is divided into multiple IHRs, where each IHR is an integrated unit of several types of DERs that are treated as a single resource with a single interconnection point to the grid.
This is feasible to do when the region of the grid representing the IHR is such that the voltage variation within each IHR is within a small $\delta$ as \cite{hosseini2022hierarchical}:
\begin{flalign}
|V_{it} - V_{jt}|<\delta,~~ \forall t, \forall b,b', \in \mathcal{B}_h, \forall h \in \mathcal{H}, \label{eq:IHR_volt}
\end{flalign}
where $\mathcal{H}$ is the set of all IHRs and $b,b', \in \mathcal{B}_h$ represent any pair of buses in IHR $h \in \mathcal{H}$.
Hence, each IHR is treated as a regular matching market with no power flow constraints, and the rest of the grid is the upstream supply source, to balance any imbalances or excesses in the IHRs.

The matching market within each IHR is an online market, with customers and RES that are characterized by uncertain arrivals and generation over time. 
In the proposed model, each IHR is endowed with an agent that can adapt its decision according to the state of the local IHR market, which includes the history of customers and renewable generation. This ensures that the decisions can be optimized with respect to the underlying state and the expected future conditioned on the current state. But, such state-dependent solutions are difficult to characterize for an online market. 
Alternatively, data-driven approaches like RL can be used to derive state-dependent solutions for systems like electricity markets, which are dynamic and uncertain. 
Given this, the IHR agents are endowed with DRL algorithms to discover state-dependent matching policies from their respective operational data. 

The design of a DRL model for a matching market has many practical limitations, such as scalability because of the size of the action space, which can grow exponentially with the number of customers, in addition to the servicing constraints that the matching outputs from the DRL model are required to satisfy. 
In this study, the scalable DRL-based solution proposed in our prior work \cite{majidi2021dynamic1} is adopted as the DRL model for each of the IHRs. 
This model is designed specifically to address the scalability and convergence of DRL applied to matching markets. 
The DRL model is discussed in the next section.

The central agent plays the role of managing the whole grid, coordinating the upstream demand of the IHRs such that the grid constraints are satisfied. 
At any moment, after the IHR-level matching decisions are determined, the IHR agents send their net active power and reactive power capacities to the central agent. 
The central agent then solves a reduced-dimension OPF problem with IHR as the nodes to compute the active and reactive power flows for the nodes such that the grid constraints (i.e., overall power flow constraints and voltage boundary limits derived from \eqref{eq:IHR_volt} are met. The voltage boundary constraints ensure that the condition in \eqref{eq:IHR_volt} is satisfied, and therefore, the matched power is delivered without much loss. The inclusion of the power flow constraints ensures that the overall flow delivers the matched power to the extent the constraints can be satisfied.
Now, a single centralized market can perform matching and be adaptable to the changing market condition, but it is typically hard to compute a solution with DRL that satisfies stability constraints, i.e., power flow constraints. 
This is the key benefit of the proposed hierarchical approach, which uses DRL to identify state-dependent policies and optimization to ensure the feasibility of the policies for grid operation.
Note that the full OPF problem is usually solved during the distribution system operation.
The proposed hierarchical model offers a matching solution that is intelligent and reduces the above operation complexity.

\subsection{Matching Market for Integrated Hybrid Resources}
\label{sec:match-prob}
This part formulates the IHR-level dynamic matching market problem for a duration of $T$, divided into time periods spaced equally at an interval $\Delta t$. 
The IHR-level dynamic matching market problem aims to match the load request of flexible and inflexible customers to maximize the social welfare in IHR $h$ subject to satisfying the supply-demand balance constraint for non-flexible loads at each time $t$ and the quality of service constraints for flexible loads arriving sequentially.
Grid power, which is denoted by $g_t \in \mathbb{R}$ and priced at $c$/kWh, together with the RES generation, $r_t \in \mathbb{R}$, are the sources of energy to supply customers in the market.

Each customer (loads) $i$ in the market is described by ${a^i,p^i,d^i,b^i}$, with $a^i \in \mathbb{N}$ as arrival time, $p^i \in \mathbb{R}$ as requested load, $d^i \in \mathbb{N}$ as the service deadline and $b^i$ as its criticality.
The customer's willingness to pay for a unit of energy can be then expressed as $c - b^i(t-a^i)$, where $b^i = \varphi c/(d^{i}-a^{i})$.
The control parameter $\varphi \in [0,1]$ allows for modeling customers with different levels of criticality.
A customer with $\varphi\!=\!1$ will only be willing to pay zero if it is served at its deadline. On the other hand, a customer with $\varphi\!=\!0$  can be served at any time without any change to the willingness to pay. This model captures a variety of customers in the market, where customers' willingness to pay can remain fixed or decay with time and at distinct rates. As the number of customers is finite in real markets, the number of customers arriving on the platform at any time $t$, $n_t \in \mathbb{N}$, is assumed to be upper bound by a constant $\overline{n}$.

Building upon the above descriptions, the market state parameters can be expressed as $z_t := [a^\top_t, p^\top_t, b^\top_t, d^\top_t, r_t]$\footnote{$[.]^\top$ denotes the transpose operation.}, where $a_t \in \mathbb{N}^{\overline{n}}$ is the vector of the arrival times of the customers which arrive at time $t$, $p_t \in \mathbb{N}^{\overline{n}}$ is the vector of their respective requested loads, $b_t \in [0,1]$ is the criticality rate of customer at time $t$, $d_t \in \mathbb{N}^{\overline{n}}$ is the vector of their respective deadlines, and $r_t \in \mathbb{R}$ is the amount of RES generation at time $t$. 
The scenario at time $t$ is given by
\begin{equation}
Z^\top_t = [z^\top_1, z^\top_2, . . ., z^\top_{t-1}, z^\top_t]. \nonumber 
\end{equation}
The probability that $z_t = z$ is given by the stochastic process modeled by $\mathbb{P}\left(z_t = z \vert Z_{t-1}\right)$. This process is not known to the market operator. 
Let $x_t := [a^\top_t, p^\top_t, b^\top_t, d^\top_t, p^\top_{u,t}, b^\top_{u,t}, r_t]$, where $p_{u,t}$ denotes the vector of the portion of the requested load that has not been served to the customers who arrived at $t$ and expressed the criticality $b^\top_{u,t}$. 
Let denote the set of all possible states at time $t$ by $\Omega_t$ and the state of the market by $X_t$. 
Then $X_t$ is given by
\begin{equation}
X^\top_t = [x^\top_1, x^\top_2, . . ., x^\top_{t-1}, x^\top_t]. \nonumber 
\end{equation}
Note that the state $X_t$ depends on the scenario $Z_t$ and the matching decisions till time $t-1$. Given that the state of the market evolves, the matching solution will have to be able to adapt to the changing market state.

Let define $A_{h,t}$ as the set of all active customers at time $t$ and IHR $h$, define $S_{h,t} = \{g,s\}$ as the set of supply types, define $p^{i}_{h,t}$ as the skipped and unsupplied load request of the customer $i$ and define $M_{h,t}(j,i,X_t) \in \mathbb{R}$ as the amount of supply of type $j$ matched to customer $i$ at time $t$ and IHR $h$, at the unit cost of $c_{h,t}$.
The matching market problem can be then stated as:
\begin{align}
&\!\!\!\!\!\!\text{sup} \sum_{t = 1}^T \sum_{i \in A_{h,t}} \sum_{j \in S_{h,t}} (\pi^{i}_{h,t} - c_{h,j})M_{h,t}(j,i,X_t), \label{eq_M:eq1} \ \text{s.t.}  \\
& \!\sum_{j \in S_{h,t}} M_{h,t}(j,i,X_t) \leq p^{i}_{h,t}, \forall h \in \mathcal{H}, \forall i \in A^h_t,  t \ne d^{i}_h,  \label{eq_M:eq2} \\
&\! \sum_{j \in S_{h,t}} M_{h,t}(j,i,X_t) = p^{i}_{h,t}, \forall h \in \mathcal{H}, \forall i \in A_{h,t}, t = d^{i}_{h},  \label{eq_M:eq3} \\
&\!\sum_{i \in A_{h,t}} M_{h,t}(\text{r},i,X_t) \leq r^{p}_{h,t},  \label{eq_M:eq4}\forall h \in \mathcal{H}, \forall t.  \\
&\! p_{h,t}^{Net} = \sum_{i \in A_{h,t}} M_{h,t}(\text{g},i,X_t),  \label{eq_M:eq5}\forall h \in \mathcal{H}, \forall t,  \
\end{align}  
where the dependency of $M_t$ on $X_t$ accounts for the dependency of the matching decision on the full state information in each IHR. 
Here, the power balance constraint for the flexible loads is given in \eqref{eq_M:eq2}.
The power balance for the non-flexible loads and the critical flexible loads at their departure ($t\!=\!d^{i}_h$) is given in \eqref{eq_M:eq3}. 
The constraint \eqref{eq_M:eq4} limits the matching power from RES to the active power output of RES $r^{p}_{h,t}$. 
Finally, the net active power flow exchanged between the IHR and upstream grid, $p_{h,t}^{Net}$, is given by \eqref{eq_M:eq5}.

The output of the above problem is a matching policy $M_{1:T}$ for the entire duration of a day. Because of the interdependence across time, the optimal policy $M_t$ is dependent on the load arrivals and RES generation for the full day. This makes the computation of the optimal policy through the above approach infeasible in real-time operation. There are also no known explicit characterizations for $M_t$. This is what makes approaches like DRL very appealing since they are general-purpose methods that can be used to discover state-dependent policies, such as $M_t$, from just operational data. Therefore, we use a DRL algorithm to compute the matching decisions. The proposed DRL model for a specific IHR is designed to output a matching decision at any point of time, depending on the market state of the IHR. The DRL framework for the IHRs matching is described in the next section. 
 
Once the matching decisions are computed by the IHRs, the agent determines the net active power flow, i.e., the active power to be taken or injected from and to the upstream grid, as well as the reactive power capacities of its zone to the central agent. 
The reactive power capacities are utilized by the central agent to adjust the nodal reactive power demands such that the constraint \eqref{eq:IHR_volt} is satisfied. 
The reactive power {capacities} for each IHR, denoted by $\underline{q}_{h,t}^{Net}$ and $\overline{q}_{h,t}^{Net}$, are obtained through \eqref{eq:Q_cap1}-\eqref{eq:Q_cap3}: 
\begin{align}
&\!\!\underline{r}^{q}_{h,t}\!\!=\!-\!\sqrt{{\overline{r}^{s}_{h,t}}^{2}\!-\!{\overline{r}^{p}_{h,t}}^{2}}~,~~\overline{r}^{q}_{h,t}\!\! = \!\!\sqrt{{\overline{r}^{s}_{h,t}}^{2}\!-\!{\overline{r}^{p}_{h,t}}^{2}}, \forall h\!\in\!\mathcal{H},\!\forall t,\label{eq:Q_cap1} \\
&\!\!\underline{q}_{h,t}^{Net}=\sum_{i \in A_{h,t}} q^{i}_{h,t}- \overline{r}^{q}_{h,t},~~\forall h \in \mathcal{H},\forall t, \label{eq:Q_cap2} \\
&\!\!\overline{q}_{h,t}^{Net}=\sum_{i \in A_{h,t}} q^{i,}_{h,t}-\underline{r}^{q}_{h,t},~~\forall h \in \mathcal{H},\forall t, \label{eq:Q_cap3} \
\end{align}
where $\underline{r}^{q}_{h,t}$, $\overline{r}^{q}_{h,t}$ are the minimum and maximum reactive power output of RES,  $\overline{r}^{s}_{h,t}$ is the available RES generation and the term $ q^{i}_{h,t}$ represents the reactive power load of the IHR, determined based on the non-flexible reactive power load and matched power to flexible loads.

\subsection{Reduced-Dimension Optimal Power Flow}
\label{sec:up-opf}
This section describes the reduced-dimension OPF problem solved by the central agent to determine the final flows for the nodes in the network to deliver the matched power to the extent it does not violate the grid constraints. The agent specifically solves a quadratic optimization model with the IHRs as the nodes, where the constraints are the power flow constraints with the active power flow demand and reactive power capacities of the IHRs, and the voltage limit constraints, defined based on \eqref{eq:IHR_volt}. The voltage limit constraints ensure that the final flows are consistent with the matched power in each of the IHRs. The central agent also curtails the active power flow demand by $p^h_C$ to the extent that the flow constraints are satisfied. 
The central agent's objective function, which is the distribution system cost, is given in \eqref{eq:C_obj}:
\begin{align}
&\min~ \big( \lambda^{RT}  P^{G} \!-\!\sum_{h = 1}^H  \lambda^{C} p_{h}^{C} \big), \label{eq:C_obj}\
\end{align}
where $P^{G}$ is the active power taken from the transmission system at the real-time market price $\lambda^{RT}$ and $p_{h}^{C}$ is the active load request curtailment with the unit cost of $\lambda^{C}$.

\subsubsection{Power Balance Constraints}
The active and reactive energy balance equations for the slack buses in the distribution system are given in \eqref{eq.Slack_bal_a}-\eqref{eq.Slack_bal_r}, where $P_{1h}$, $Q_{1h}$ are the active and reactive power flows from the substation bus to the IHR $h$, $V_{1}^{sq}$ is the squared voltage on the substation bus, $g_1$, $b_1$ are the shunt conductance and susceptance at the substation bus and $\mathcal{L}$ is the set of lines in the distribution grid.
\begin{align}
&P^{G} = \sum\limits_{1h \in \mathcal{L}} P_{1h}  + {g_1}V_{1}^{sq}\label{eq.Slack_bal_a}, \\
&Q^{G} = \sum\limits_{1h \in \mathcal{L}} Q_{1h}  + {b_1}V_{1}^{sq}\label{eq.Slack_bal_r}.\
\end{align}

The energy balance constraints for the IHR nodes are presented in \eqref{eq.No_slack_bal_a}-\eqref{eq.No_slack_bal_r}, where $p_{h}^{Net}$ is the net active power flow submitted by the IHR, $P_{hh''}$, $P_{h'h}$ and $Q_{hh''}$, $Q_{h'h}$ are the active and reactive power flows in lines $hh''$ and $h'h$, $V_{h}^{sq}$ is the squared voltage on IHR node $h$, $I_{h'h}^{sq}$ is the squared current flow in line $h'h$, and $r_{h'h}$, $x_{h'h}$ are the resistance and reactance of the line $h'h$ and $g_h$, $b_h$ are the shunt conductance and susceptance at the IHR node $h$. In the active power balance constraint, the curtailment $p_{h}^{C}$ ensures that the final net active power flow is consistent with the power flow constraints. Here, the curtailment is only made to the extent that the constraints are satisfied. The reactive power flow of the IHRs, denoted by $q_{h}^{Net}$, is limited to the reactive power capacities of IHRs in \eqref{eq.q_min_max}.
\begin{align}
&\!\!\!\!P_{hh''}\!+\!p_{h}^{Net}\!-\!p_{h}^{C}\!=\!\!\!\!\sum\limits_{h'h \in \mathcal{L}}\!\! {\left( {P_{h'h}\!\!-\!\! {r_{h'h}}I_{h'h}^{sq}}\right)\!+\!{g_h}V_{h}^{sq}}\label{eq.No_slack_bal_a},\forall h\!\in\!\mathcal{B},\!\\
&\!\!\!\!Q_{hh''}\!+q_{h}^{Net}\!=\!\sum\limits_{h'h \in \mathcal{L}}\!\!\!{\left( {Q_{h'h}\!- {x_{h'h}}I_{h'h}^{sq}} \right)\!+{b_h}V_{h}^{sq}} \label{eq.No_slack_bal_r},\forall h \in \mathcal{B},\!\!\\
& \underline{q}_{h}^{Net} \le q_{h}^{Net} \le \overline{q}_{h}^{Net}, \forall h. \label{eq.q_min_max}\
\end{align}

\subsubsection{Voltage and Power Flow Limits}
The voltage drop across the grid is given by \eqref{eq.V_drop}. The limits on the squared voltage level and the limits on the current flow are given in Eq. \eqref{eq.V_lim} and Eq.  \eqref{eq.I_lim}, where $\underline{V}_h^{sq}$, $\overline{V}_h^{sq}$ are the minimum and maximum squared voltage boundaries, defined based on the nominal node voltage and $\delta$ in \eqref{eq:IHR_volt} and $\overline{I}_{h'h}^{sq}$ is the squared current flow limit.
Finally, the complex power flow constraint is given in \eqref{eq.p_lim}.
\begin{align}
&V_{h}^{sq} - V_{h'}^{sq} = - 2\left( {{r_{h'h}}P_{h'h} + {x_{h'h}}Q_{h'h}} \right) \nonumber \\
&~~~~~~~~+\left( {r_{h'h}^2 + x_{h'h}^2} \right)I_{h'h}^{sq}\label{eq.V_drop},~~~~~\forall (h'h) \in \mathcal{L},\\
&\underline{V}_h^{sq} \le V_{h}^{sq} \le \overline{V}_h^{sq}\label{eq.V_lim},~~~~~\forall h \in \mathcal{B},\\
&I_{h'h}^{sq} \le \overline{I}_{h'h}^{sq}\label{eq.I_lim},~~~~~\forall (h'h) \in \mathcal{L},\\
&V_{h,t}^{sq}I_{h'h}^{sq} \ge P_{h'h}^2 + Q_{h'h}^2\label{eq.p_lim},~~~~~\forall (h'h) \in \mathcal{L}.\
\end{align}

Any feasible solution to the online OPF problem in \eqref{eq:C_obj}-\eqref{eq.p_lim} ensures the matched power in each of the IHRs is delivered to the extent the flow and voltage constraints are met. 
In case a solution is feasible without any curtailment, then the matched power will be delivered to the customers.

\section{Deep Reinforcement Learning for IHRs} \label{DRL}
In the proposed hierarchical framework, {the dynamic matching of IHRs is formulated as a Markov Decision Process (MDP) with the following components: (i) state that includes real-time energy price, RES generation, supplied energy, active customers, and their deadline and dynamic willingness to pay for a unit of energy, (ii) action represented by the matching policy that is trainable and is taken by each IHR agent to determine the probability distribution over the set of matching decisions for the flexible loads and RES, and (iii) reward that is reflected as the social welfare in IHR dynamic matching market}. 
A policy gradient RL algorithm is applied to train the matching policy given the load and generation data for multiple instances of the market. 
This algorithm does not require supervision or expert knowledge as it measures its own performance for the training process. 
The following subsections briefly discuss the matching policy's structure for an IHR and then the learning algorithm. 
Note that the expectation with respect to all sources of randomness is denoted by $\mathbb{E}[.]$.
It is implicit that all the descriptions in this section are confined to a single IHR.

\subsection{General Discrete Matching Policy}
Each IHR agent in the proposed study learns an online matching policy given by $\chi = \{\chi_1, \chi_2, \chi_3,...,\chi_T\}$.
The discrete matching policy for time $t$, denoted by $\chi_t$, indicates whether a customer is to be matched to a supply or not, regardless of the amount of matching.
Let define $\mathcal{M}_t$ as the space of discrete matching at time $t$.
Each component in this set $m \in \mathcal{M}_t$, is a feasible discrete matching that specifies whether a customer is matched or not (i.e., $m_{i,k} \in \{0,1\}$ with one indicating ``matched" and zero indicating ``not matched").
Hence, the general matching policy $\chi_t$ can be given by:
\begin{equation}
\chi_t : \Omega_t \rightarrow \mathcal{M}_t. \nonumber
\end{equation}

Aside from the fact that the matching problem is an online decision-making procedure with a future ridden with uncertainties, there are still several general challenges from an RL point of view. 
Firstly, the action space of the matching problem is large and specifically exponential in the number of customers. For example, if there are $m$ supplies and $n$ customers, then there are $m^n$ ways of matching; thus, it is exponential in the number of active customers. Secondly, not all the actions from this space are feasible, as supply unavailability might limit the matching decisions. There might also be some restrictions enforced by the customers' servicing constraints. Hence, some actions are infeasible, and their infeasibility is state-dependent. Thirdly, RL algorithms can converge to a local optimum, a general challenge that applies to the matching problem. Therefore, the goal is to develop a framework based on RL that is simple and efficient to learn, simultaneously satisfies the action constraints, and can converge to a good solution. The proposed framework in this paper simplifies the output of the policy to be trained by RL to just ``match" or ``not to match" for each active customer, regardless of the supply type and action feasibility. 
Thus, the action space of the output of the component that is trained is linear in the number of active consumers. Further details regarding the proposed matching policy are given below.

\subsection{Proposed Matching Policy}
The proposed discrete matching policy is characterized by a learnable and fixed component \cite{majidi2021dynamic1}.
The first component, denoted by $\mu_t$, determines the probability of matching customers, and the latter makes sure that the customers are matched before their deadline, (i.e., avoiding infeasible actions during the training process and real-time implementation).
Let $\mathcal{P}_{\mathcal{M}_t}$ be the space of probability measures over the set $\mathcal{M}_t$. 
Then, the policy $\mu_t$ can be defined:
\begin{equation}
\mu_t : \Omega_t \rightarrow \mathcal{P}_{\mathcal{M}_t}. \nonumber
\end{equation}
Let $m_t \in \mathcal{M}_t$ be given by $m_{t} \sim \mu_t$. 
The component of $m_t$ corresponding to the customer $i$ is defined by $m_{i,t}$, where $m_{i,t} \in \{0, 1\}$. The output $m_i$ is input to a second function, $\varphi$. The function $\varphi$ matches the customers with $m_{i,t} = 1$ to the available RES in each IHR. When total matching implied by the discrete matching is in excess of the RES, the remaining customers with $m_{i,t} = 1$ are matched to the grid supply. 
When total matching implied by the discrete matching is less than the available RES, the excess RES generation is assigned to the remaining customers.
Denote the component of $\varphi$ that specifies whether customer $i$ is matched to supply type $j$ by $\varphi_{j,i}$. 
The output $\varphi_{j,i}$ is input to a third function, $\nu$, that overturns the matching decision for the customers with an immediate deadline and ensures that the flexible customers in IHRs are served by their deadline:
\begin{equation}
\nu_{j,i} = \left\{ \begin{array}{cc} 1 & \text{if} \ d^i = t, \ \text{$i$ is active}, \ \varphi_{s,i} = 0 ~ \forall s \\ & \text{and} \ j = \text{g}, \\
\varphi_{j,i} & \ \text{otherwise}. \end{array} \right. \nonumber
\end{equation}
Thus, the overall discrete matching policy for time $t$, $\chi_t$, is given by:
\begin{equation}
\chi_t = \nu \circ \varphi \circ m_t, m_t \sim \mu_t.
\label{eq:matchingalg}
\end{equation}
The proposed policy is parameterized by $\theta_t$, where the parameterization is denoted by $\mu_t(.;\theta_t)$. 
The learning algorithm presented next, uses the observations from load and generation data of the IHR and trains $\theta_t$ for every time step $t$ by evaluating its own performance.

\subsection{Policy Gradient Learning Algorithm}
This part describes the proposed policy gradient learning algorithm. $\mathbb{E}_{X_t \sim \mathbb{P}_t(.)}$ is used as a shorthand for expectation over $X_t \sim \mathbb{P}(. \vert X_{t-1}, \chi_{t-1})$, where $\mathbb{P}(. \vert X_{t-1}, \chi_{t-1})$ denotes the transition probability from state $X_{t-1}$ under the policy $\chi_{t-1}$. 
Let $m_{t:T} = \{m_t, m_{l+1},...,m_T\}$ and $\mu_{l:T} = \{\mu_t, \mu_{l+1},..., \mu_T\}$. Denoting $\mathbb{E}_{m \sim \mu}$ as a shortened form of $\mathbb{E}_{m_{t+1:T}\sim \mu_{t+1:T}}$, the market welfare can be defined as:
\begin{align}
&\!\!\!V^\chi_{t+}(X_{t+1})\!:=\!\mathbb{E}_{m \sim \mu} \sum_{l=t+1}^T\sum_j \sum_{k \in A_t} (\pi^i_l - c_j) \chi_{t,j,i}(X_t) . \label{Gra_pol_1}\
\end{align}
Let:
\begin{align}
& v^\chi_t := \sum_j \sum_{i \in A_t} (\pi^i_l - c_j) \chi_{t,j,i}(X_t) ,\label{Gra_pol_2}\\
& V^\chi_t(X_t\vert m_t) := v^\chi_t + \mathbb{E}_{X_{t+1} \sim \mathbb{P}_{t+1}(.)}\left[V^\chi_{t+}(X_{t+1})\right].\label{Gra_pol_3}\ 
\end{align}

Then, from the definitions of $V_\chi$ and $V^\chi_t(X_t)$, the gradient of the value function with respect to the policy parameter $\theta_t$ can be calculated as follows:
\begin{align}
&\frac{\partial V_\chi}{\partial\theta_t} = \mathbb{E}_{X_t}\left[\frac{\partial V^\chi_t(X_t)}{\partial\theta_t}\right],\label{Gra_pol_4}\\
& \frac{\partial V_\chi}{\partial\theta_t} = \mathbb{E}_{X_t}\sum_{m_t \in \mathcal{H}_t}\frac{\partial\mu_{t}(m_t;\theta_t)}{\partial \theta_t} \left[v^\chi_t \right. \nonumber\\
&~~~~~~~~~~~~~~~~~\left. + \mathbb{E}_{X_{t+1} \sim \mathbb{P}_{t+1}(.)} V^\chi_{t+}(X_{t+1})\right].\label{Gra_pol_5}\
\end{align}

The gradient of the value function with respect to the policy parameter derived in \eqref{Gra_pol_4}-\eqref{Gra_pol_5} can be written as follows:
\begin{align}
&\frac{\partial V_\chi}{\partial\theta_t} = \mathbb{E}_{X_t, m_t \sim \mu_t} \left[\frac{\partial \log\mu_{t}(m_t;\theta_t)}{\partial \theta_t} V^\chi_t(X_t\vert m_t)\right], 
\label{Gra_pol_6}\
\end{align}
where an unbiased estimate of this relationship can defined as follows:
\begin{align}
&\delta^\theta_t = \left[\frac{\partial \log\mu_{t}(m_t;\theta_t)}{\partial \theta_t} V^\chi_t(X_t\vert m_t)\right].\label{Gra_pol_7}\
\end{align}

Since the term $V^\chi_t(X_t\vert m_t)$ is unknown, the gradient in \eqref{Gra_pol_7} is not computable.
Therefore, this term is replaced with the social welfare from $t$ to $T$ for a sample epoch under policy $\chi$:
\begin{align}
&\delta^\theta_{t,r} = \frac{\partial \log\mu_{t}(m_t;\theta_t)}{\partial \theta_t} \left[\sum_{l=t}^T v^\chi_l \right],\label{Gra_pol_8}\
\end{align}
where the gradient is computable using the data from a sample epoch $\left(Z = \{Z_1,Z_2,...,Z_T\}\right)$ and matching decisions under the policy $\chi$ for the same epoch. Furthermore, the gradient estimate is unbiased as $\frac{\partial V_\chi}{\partial\theta_t} = \mathbb{E}[\delta^\theta_{t,r}]$.
The vanilla policy gradient learning algorithm learns the policy parameter $\theta_t$ for each time step using the following stochastic gradient ascent algorithm:
\begin{align}
\theta_{t+1} \leftarrow \theta_t + \gamma_\theta \delta^\theta_{t,r}, \label{Gra_pol_9}\
\end{align}
where $\theta_t$ is updated using the computed gradient $\delta^\theta_{t,r}$ for multiple sample epochs in every update step. 

In addition to the vanilla policy gradient learning algorithm described above, the actor-critic algorithm AC$-k$ is also proposed for the dynamic matching of IHRs. 
This algorithm learns both the matching policy $\mu$ and an approximation of value function $V^\chi_t(X_t)$. 
This function, which is also called the critic function, is parameterized by $\phi_k$ and expressed by $V^\chi_k(X_k;\phi_k)$. 
Hence, the approximate policy gradient for the actor-critic algorithm AC$-k$ can be defined as follows:
\begin{align}
&\!\!\!\!\delta^\theta_{t,k} = \!\frac{\partial \log\mu_{t}(m_t;\theta_t)}{\partial \theta_t} \!\!\left[\sum_{l=t}^{t+k-1} v^\chi_l + V^\chi_{t+k}(X_{t+k};\phi_{t+k})\!\right]\!, \label{Gra_pol_10}\
\end{align}
where the policy parameters are learned using the following stochastic gradient ascent algorithm in \eqref{Gra_pol_11} and similarly, the parameter $\phi_k$ of the critic function is learned by stochastic gradient descent for its least-squares error in \eqref{Gra_pol_12}.
\begin{align}
&\theta_{t+1} \leftarrow \theta_t + \gamma_\theta \delta^\theta_{t,k},\label{Gra_pol_11}\\
&\phi_{k+1} \leftarrow \phi_k - \gamma_\phi \left(V^\chi_{k}(.;\phi_k) - \sum_{l=k}^T v^\chi_l\right).\label{Gra_pol_12}\
\end{align}
Further details regarding the actor-critic algorithm are given in Algorithm \ref{alg:policygradient}, where the ADAM gradient algorithm of the gradient updates in \eqref{Gra_pol_11} and \eqref{Gra_pol_12} is implemented.

\begin{algorithm}[!th]
%\floatname{algorithm}{}
%\label{alg:rh-Alg}
\setlength{\leftmargini}{0.2in}
\begin{enumerate}
\item Initialize $\mathcal{D} = \varnothing$, $j = 0$
\item Initialize $\theta_k \ \forall \ k \in [1,...,T]$. $N$: number of epochs
\item for $i = 1:N$

\begin{enumerate}
\item j = j + 1
\item Set $D_i = \{\{X_k,m_k,v^\chi_k\}\ \forall \ k \in [1,...,T]\}$
\item Include $D_i$ into $\mathcal{D}$
\item if j == M \\
Update $\theta_k$ by ADAM of Eq. \eqref{Gra_pol_11} using $\mathcal{D}$ \\
Update $\phi_k$ by ADAM of Eq. \eqref{Gra_pol_12} using $\mathcal{D}$ \\
j = 0; $\mathcal{D} = \varnothing$ \\
end
\end{enumerate}
end
\item end
\end{enumerate} 
\caption{\bf Actor-Critic (AC$-k$) Policy Gradient Learning Algorithm for an IHR}
\label{alg:policygradient}
\end{algorithm}

The matching policy in the proposed study is trained with the Temporal Convolution Network (TCN), denoted by TCN$_\mu$.
Let $\tilde{X}_t$ denote the input sequence to the TCN at each time step, where $\tilde{X}^\top_t = [\tilde{x}^\top_1, \tilde{x}^\top_2, \tilde{x}^\top_3,...,\tilde{x}^\top_t]$, and $\tilde{x}_t =[a^\top_t, p^\top_t, b^\top_t, d^\top_t, p^\top_{u,t}, b^\top_{u,t}, r_t]$.
The vector of matching probabilities $P^m_\mu \in [0,1]^{\overline{n}\times T}$ as the output of TCN can be determined as $P^m_\mu = \text{TCN}_\mu(\tilde{X}_t)$.  
The output of TCN in the proposed study is fixed and capped at the maximum number of active customers at any time, $\overline{n}\times T$.
Let $P^m_{\mu,i}$ denote the probability of matching for the active customer $i$ at time step $t$. 
Hence, the distribution $\mu_t$ is constructed as:
\begin{equation}
\mathbb{P}(m_{t,i} = 1) = P^m_{\mu,i}, \ \mathbb{P}(m_{t,i} = 0) = 1 - P^m_{\mu,i}. \nonumber
\end{equation}

\section{Simulations and Results} \label{Simulations}
\begin{figure}[ht!]
\centering{
\includegraphics[width=1\linewidth]{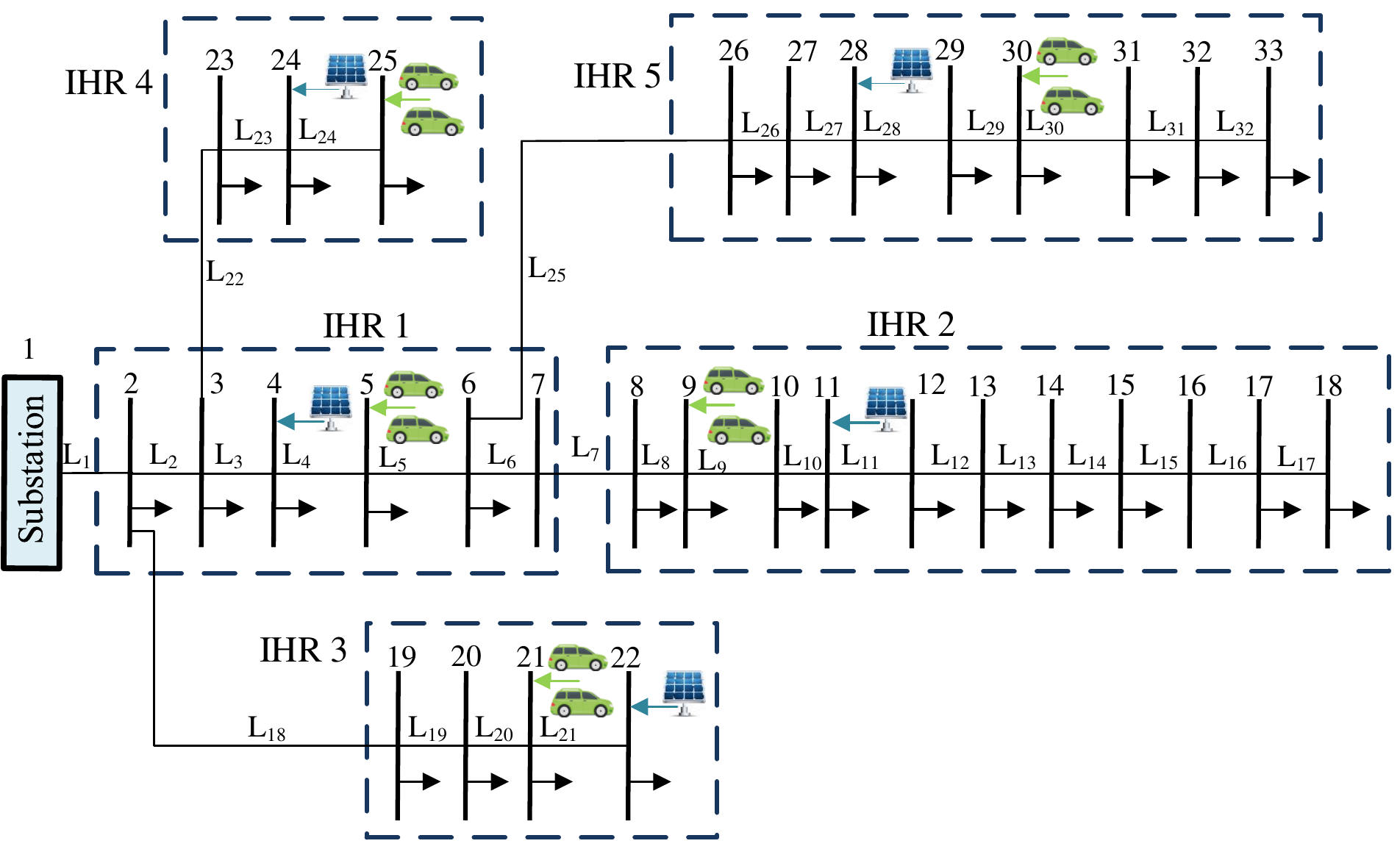}}
\caption{Structure of the 33-bus power distribution system, divided into 5 integrated hybrid resources.} 
\label{fig:PDS}
\end{figure}

The proposed hierarchical matching framework is implemented on the IEEE 33-bus test distribution system using the 30-minute real-time California Independent System Operator (CAISO) load and solar generation data from January 1, 2021, to September 28, 2021. 
The IHR nodes are determined based on similar loading features (i.e., voltage profiles) determined from running an OPF problem to minimize the power loss for the peak distribution system load.
{Hence, the distribution system is divided into 5 IHRs, each consisting of a learning agent to control and match DERs with flexible loads such that \eqref{eq:IHR_volt} is satisfied.}
The structure of the distribution system with IHRs is shown in Fig. \ref{fig:PDS}, where the electric vehicle (EV) charging stations supply charging requests of 6.6 kWh to 24 EVs in IHR 1, 30 EVs in IHR 2, 8 EVs in IHR 3 and 30 EVs in each one of IHRs 4 and 5.
The CAISO solar power data is scaled to the inverter's nominal capacity of 105 kW in IHR 1, 150 kW in IHR 2, 45 kW in IHR 3, and 150 kW in each one of IHRs 4 and 5.
The distribution system active and reactive loads are scaled to 50 \% of their nominal rates, 3715 kW and 2300 kVAr, respectively. The electricity tariff is assumed to be 120 \$/MWh, and the curtailment penalty for the central agent is assumed to be 500 \$/MWh. 
To validate the efficiency of the proposed hierarchical framework, the following scenarios are considered:
\begin{itemize}
    \item \textit{Scenario 1}: This is a scenario where the EVs are characterized by earlier arrival times and longer departure times. In this scenario, waiting to match will fetch higher welfare. This scenario tests the capability of the IHR agents to learn to let the customers wait in the market and not match them immediately upon their arrival.  
    \item \textit{Scenario 2}:  This is a scenario where the EVs are characterized by moderate arrival and longer departure times. In this scenario, waiting may not result in improved welfare. Here, a strategy that partially waits and partially matches upon arrival might be needed. This scenario tests the capability of the algorithm to learn such hybrid strategies.
\end{itemize}

For illustration, two matching algorithms are considered: one is the Learning Algorithm (LA) described in Section \ref{DRL}, and the other is the standard matching heuristic, Matching on Arrival (MA). These matching algorithms are implemented in scenarios 1 and 2 under the following market models:
\begin{itemize}
    \item \textit{Centralized Model}: In this model, a single agent manages the matching of the whole distribution system.   
    \item \textit{Decentralized Model}: In this model, the distribution system is divided into multiple IHRs, as described earlier, with each IHR employing a separate matching algorithm. 
    The central agent solves the reduced-dimension OPF model described earlier to meet the respective IHRs flow requirements and ensure that the grid constraints are met. This model with the learning algorithm is the proposed hierarchical framework.
\end{itemize}

The best hyper-parameters for the TCN model were identified to be 3 for the number of blocks, 4 for the number of filters, 3 for the filter size, 0.1 for the dropout factor, and 4 for the dilation factor. 
Sigmoid function is utilized as the activation function for each output of TCN and the following values are considered as the parameters of the ADAM algorithm: ${\alpha}=[0.25,0.99]$, ${\beta}_1=0.9$, ${\beta}_2=0.999$, ${\epsilon}=10^{-8}$, where $\alpha$ is the learning rate and ${\beta}_1$, ${\beta}_2$ are exponential decay rates for the moment estimates. 
The best batch size for the LA is 20.

\subsection{Numerical Results}
The average social welfare achieved in scenarios 1 and 2 for the centralized and decentralized models is summarized in Table \ref{tab:Numerical_results}.
In scenario 1, it can be seen that the MA algorithm achieves a welfare of 17.98\$ and 16.17\$ in both the centralized and decentralized models, while the LA algorithm achieves a higher welfare of 218.59\$ and 232.7\$ in the centralized and decentralized models, respectively. This shows that the LA algorithm leverages the flexibility better to match the excess RES during the middle of the day. In scenario 2, the results reveal that the optimal matching policy is not to match all the loads on their arrival but to only match the critical ones on their arrival, so as to efficiently utilize the RES that is available during the middle of the day. In this scenario, the LA algorithm is the top-performing, achieving a social welfare of 866.33\$ and 893.52\$ in the centralized and decentralized models, followed by the MA algorithm, which achieves a social welfare of 742.75\$ and 808.2\$ in the centralized and decentralized models, respectively.

Comparing the performance of LA and MA algorithms in the centralized and decentralized models, it can be seen that the decentralized model with the learning algorithm is the best performing, substantiating the efficacy of our approach. Table \ref{tab:Numerical_results_Dece} summarizes the social welfare achieved by LA and MA in the decentralized matching model.
Comparing the social welfare, it can be found that the LA algorithm outperforms the MA algorithm in each of the IHRs, showing the superiority of the learning-based approach. 

\begin{table}
\centering
\caption{Average Social Welfare in Centralized and Decentralized Models (\$)}
\label{tab:Numerical_results}
\resizebox{\columnwidth}{!}{%
\begin{tabular}{cccc}
\hline
\textbf{Algorithm} & \textbf{Scenario} & \textbf{Centralized Model} & \textbf{Decentralized Model} \\ \hline
\multirow{3}{*}{\textbf{LA}} & \textbf{Scenario1} & $218.59$ & $232.70$ \\ \cline{2-4} 
 & \textbf{Scenario2} & $866.33$ & $893.52$ \\  \cline{2-4}
 & \textbf{Average} & $\mathbf{542.46}$ & $\mathbf{563.11}$ \\\hline
\multirow{3}{*}{\textbf{MA}} & \multicolumn{1}{l}{\textbf{Scenario1}} & ${17.98}$ & ${16.17}$ \\ \cline{2-4} 
 & \textbf{Scenario2} & $742.75$ & $808.20$ \\ \cline{2-4}
 & \textbf{Average} & $\mathbf{380.365}$ & $\mathbf{412.185}$ \\\hline
\end{tabular}
}
\end{table}

\begin{table}
\centering
\caption{Average Social Welfare in the Decentralized Model (\$)}
\label{tab:Numerical_results_Dece}
\resizebox{\columnwidth}{!}{%
\begin{tabular}{ccccccc}
\hline
\textbf{Algorithm} & \textbf{Model} & \textbf{IHR1} & \textbf{IHR2} & \textbf{IHR3} & \textbf{IHR4} & \textbf{IHR5}\\ \hline
\multirow{3}{*}{\textbf{LA}} & \textbf{Scenario1} & $41.65$ & $55.06$ & $12.3$ & $60.02$ & $63.67$\\ \cline{2-7} 
 & \textbf{Scenario2} & $184.15$ & $216.67$ & $49.47$ & $221.36$ & $221.87$ \\ \cline{2-7} 
& \textbf{Average} & $\mathbf{112.9}$ & $\mathbf{135.865}$ & $\mathbf{30.885}$ & $\mathbf{140.69}$ & $\mathbf{142.77}$ \\ \hline
\multirow{3}{*}{\textbf{MA}} & \textbf{Scenario1} & $3.06$ & $3.89$ & $1.02$ & $4.164$ & $4.04$\\ \cline{2-7} 
 & \textbf{Scenario2} & $163.08$ & $202.4$ & $43.53$ & $199.46$ & $199.73$ \\ \cline{2-7}
 & \textbf{Average} & $\mathbf{83.07}$ & $\mathbf{103.145}$ & $\mathbf{22.275}$ & $\mathbf{101.81}$ & $\mathbf{101.885}$ \\ \hline
\end{tabular}
}
\end{table}

\subsection{Matching Market Analysis}
This section analyzes the performance of the matching algorithms under the centralized and decentralized matching markets in scenarios $1$ and $2$.

\subsubsection{Scenario $1$: EVs with Earlier Arrival and Longer Departure Times}
\begin{figure}[ht!]
\centering{
\includegraphics[width=1\linewidth]{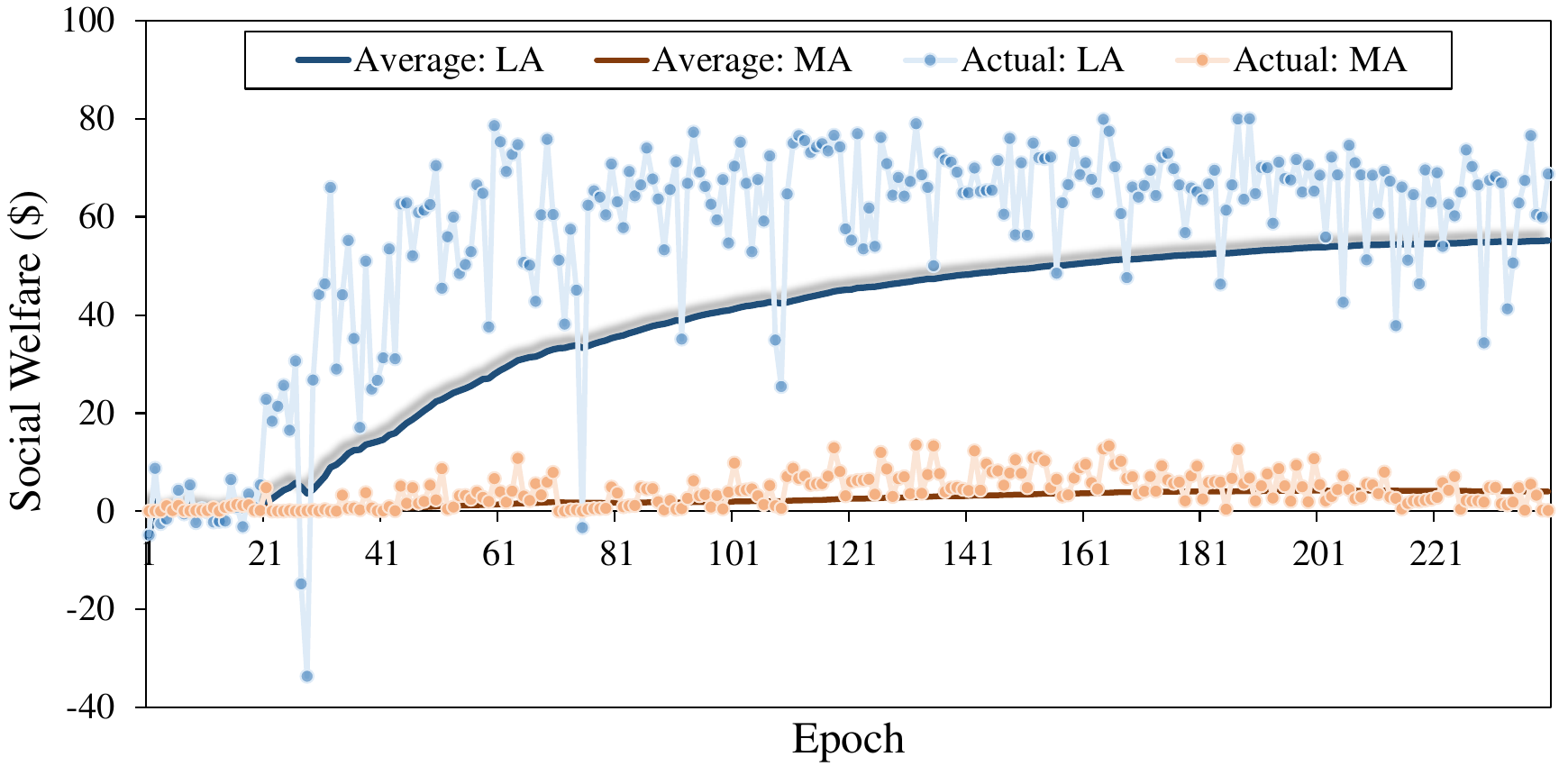}}
\caption{Average and actual social welfare of LA and MA for IHR $2$ under scenario $1$.} 
\label{fig:Agent_IHR2_sc1}
\end{figure}

In this scenario, the flexible loads are characterized by earlier arrival and longer departure (deadline) times, and the RES generation is available during the middle of the day. Thus, the market operator (agent) can queue the load requests and match them to the RES available during the middle of the day. 
The training curve that reflects the social welfare achieved by the LA and MA of IHR $2$ in this scenario is shown in Fig. \ref{fig:Agent_IHR2_sc1}. The results clearly show that MA fails to wait to avail the RES during the middle of the day, and instead matches the loads to the grid supply. On the contrary, the LA learns to queue the non-critical loads and shift them to the periods where RES generation is available.

\begin{figure}[ht!]
\centering{
\includegraphics[width=1\linewidth]{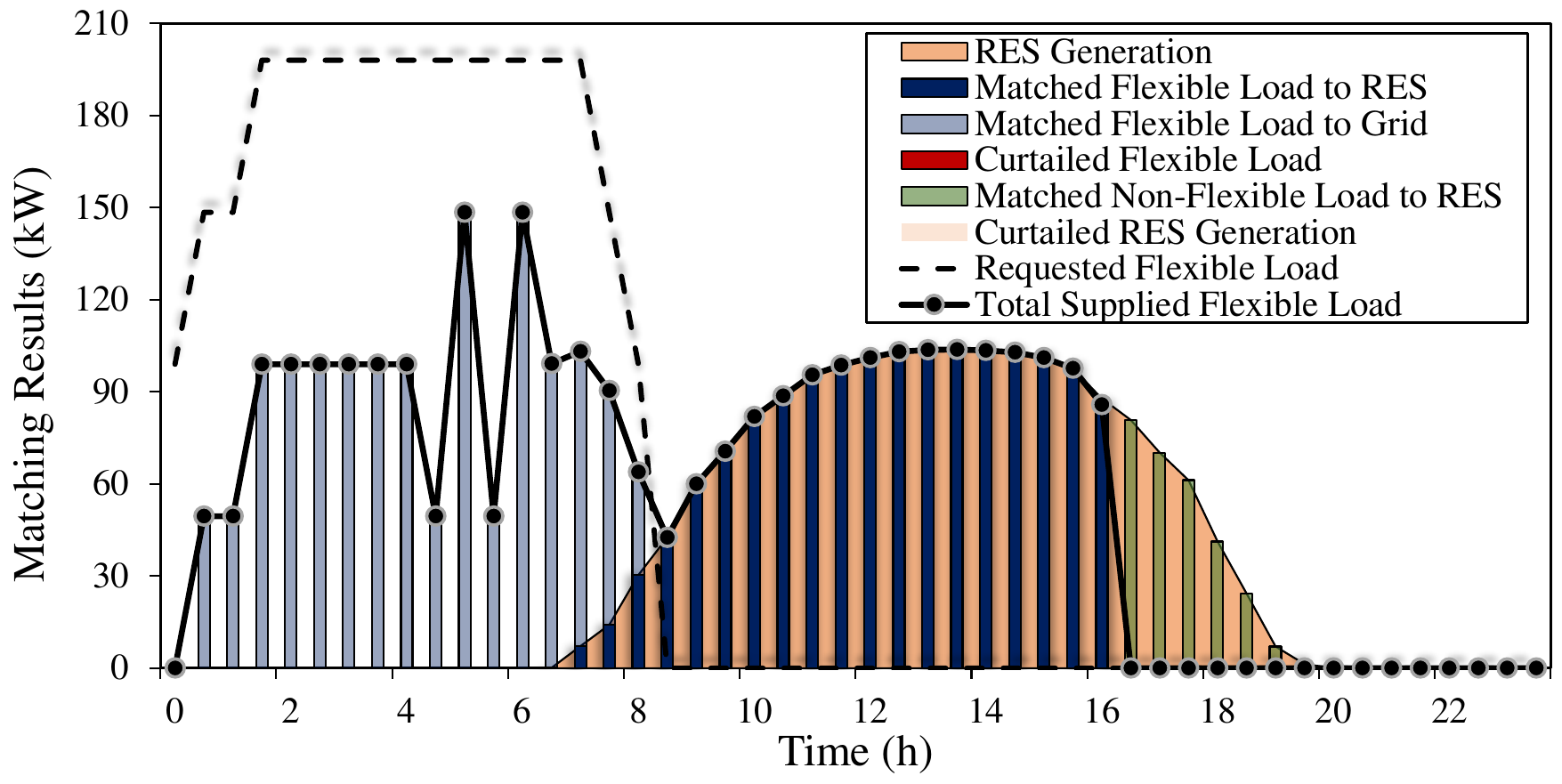}}
\caption{Matching market results for LA of IHR $2$ for a representative epoch of scenario $1$.} 
\label{fig:Results_IHR2}
\end{figure}

Figure \ref{fig:Results_IHR2} shows the matching market results for the LA of IHR $2$ for a representative epoch of scenario $1$.
In Fig. \ref{fig:Results_IHR2}, the initial load request of critical loads is supplied using the grid power, while a significant portion of non-critical loads is shifted to the middle of the day and matched to the RES generation, indicating the efficacy of the fixed and trainable components of the matching policy to match flexible loads with RES, while satisfying the quality of service constraints of the loads.
This is evident in Fig. \ref{fig:Results_IHR2}, where all the requested load is supplied without any curtailment, and the RES generation is efficiently allocated to supply the queued flexible loads and the non-flexible loads when the flexible loads are unavailable. 

\begin{figure}[ht!]
\centering{
\includegraphics[width=1\linewidth]{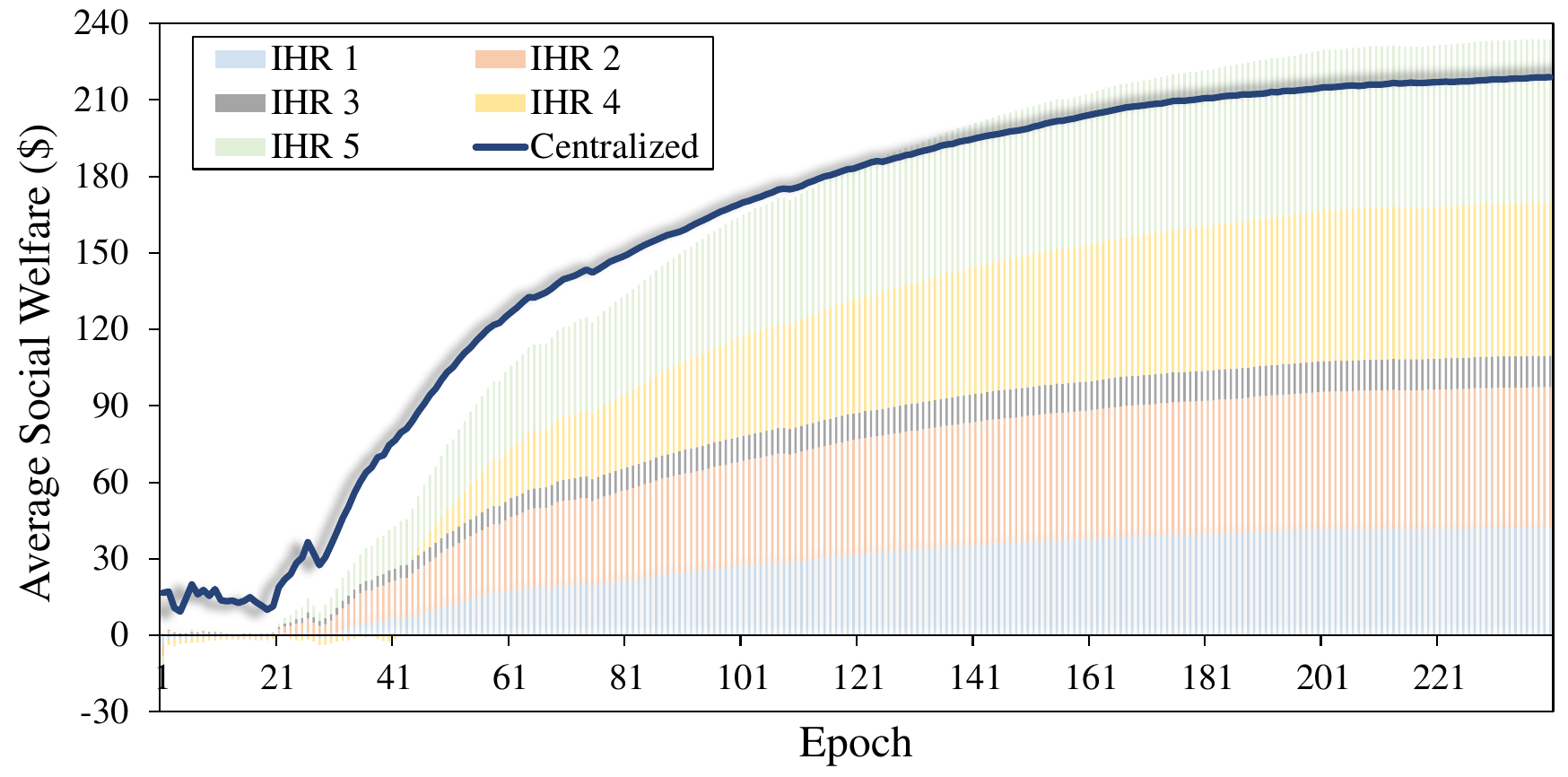}}
\caption{Average social welfare of decentralized and centralized models with LA in scenario $1$.} 
\label{fig:Results_Cetralized_LA}
\end{figure}

The training curves for learning algorithms in decentralized and centralized models are compared in Fig. \ref{fig:Results_Cetralized_LA}.
As shown, the centralized model obtains higher welfare in the initial epochs, but the decentralized model achieves a higher average social welfare with experience.

\subsubsection{Scenario $2$: EVs with Moderate Arrival and Longer Departure Times}
\begin{figure}[ht!]
\centering{
\includegraphics[width=1\linewidth]{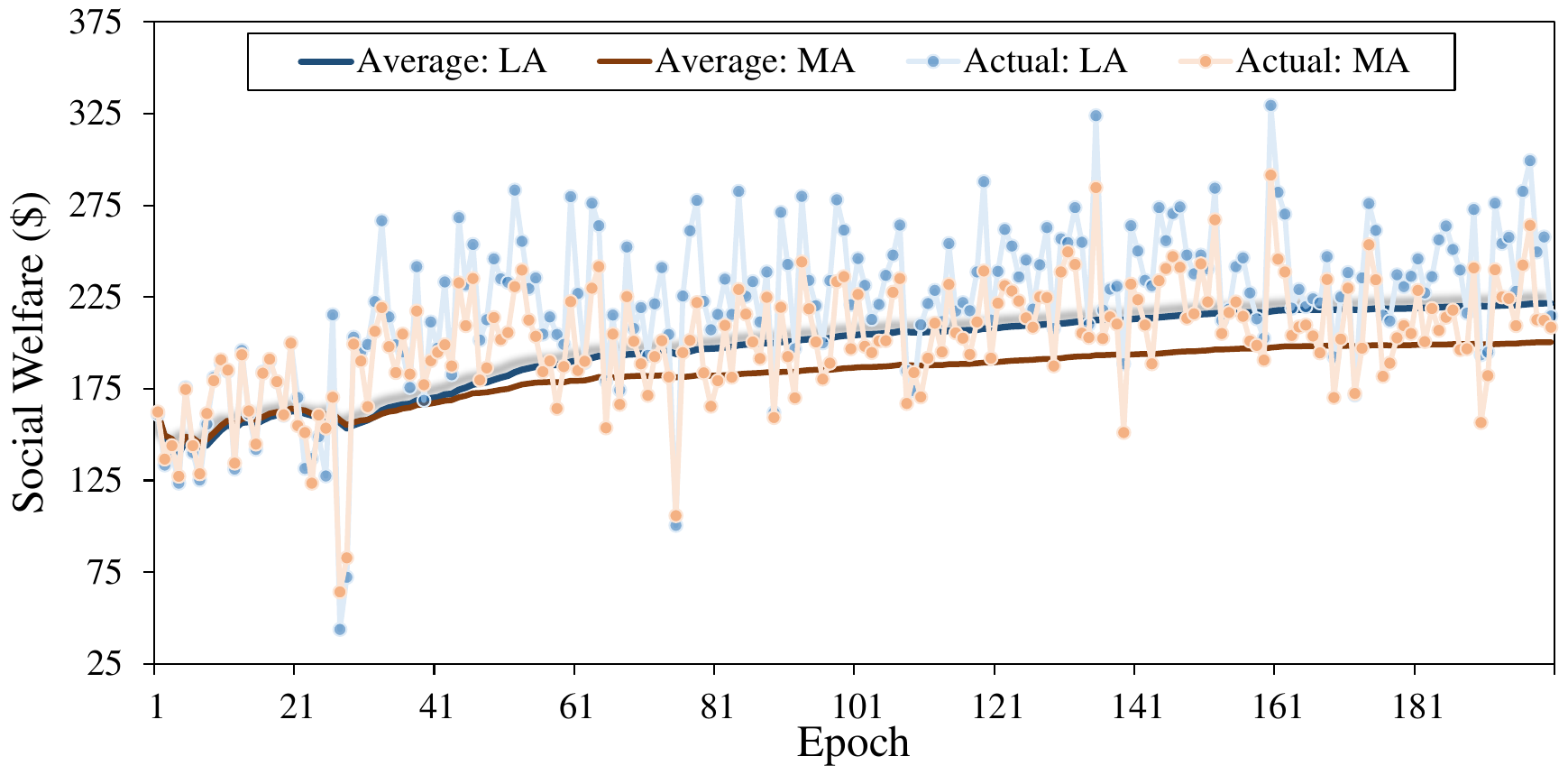}}
\caption{Average and actual social welfare of LA and MA of IHR $5$ in scenario $2$.}
\label{fig:Agent_IHR5_sc2}
\end{figure}

\begin{figure}[ht!]
\centering{
\includegraphics[width=1\linewidth]{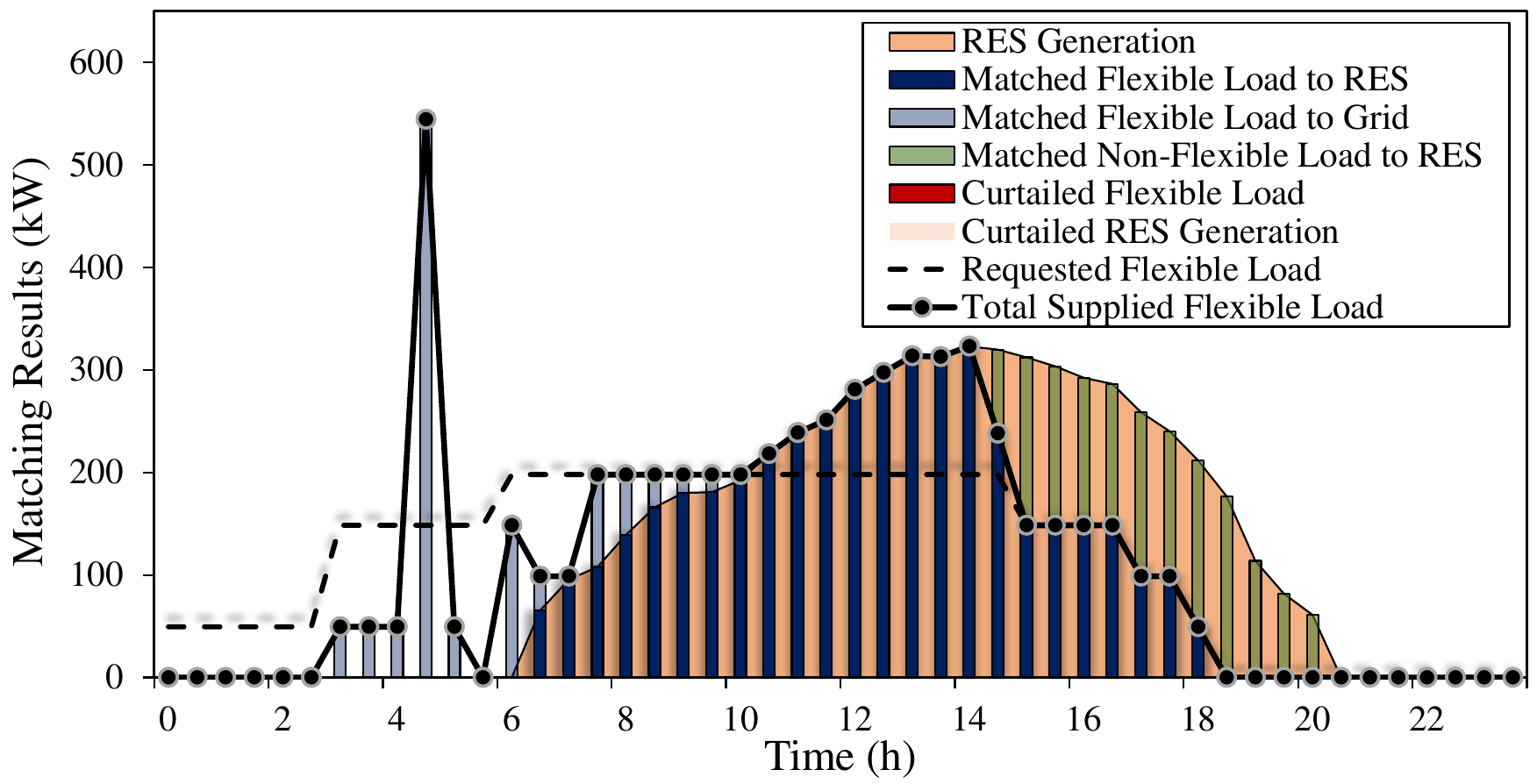}}
\caption{Matching results for LA of IHR $5$ for a representative epoch in scenario $2$.} 
\label{fig:Results_IHR5}
\end{figure}

In this scenario, the flexible loads are characterized by moderate arrival and longer departure times, and the RES generation is available during the middle of the day.
The social welfare achieved by the LA and MA of IHR $5$ in this scenario is shown in Fig. \ref{fig:Agent_IHR5_sc2}.
The results show that both the LA and MA achieve a similar performance. However, as shown, the LA is superior to the MA in that it doesn't match all the flexible loads on their arrival.
This is evident in Fig. \ref{fig:Results_IHR5}, where a portion of the load request is shifted from their arrival and matched to the RES generation during the middle of the day. As in the former scenario, the LA is able to meet the servicing constraints of the loads and utilize the RES generation, while ensuring that the outcomes are economically efficient.

\subsection{Distribution System Constraints}
\begin{figure}[ht!]
\centering{
\includegraphics[width=1\linewidth]{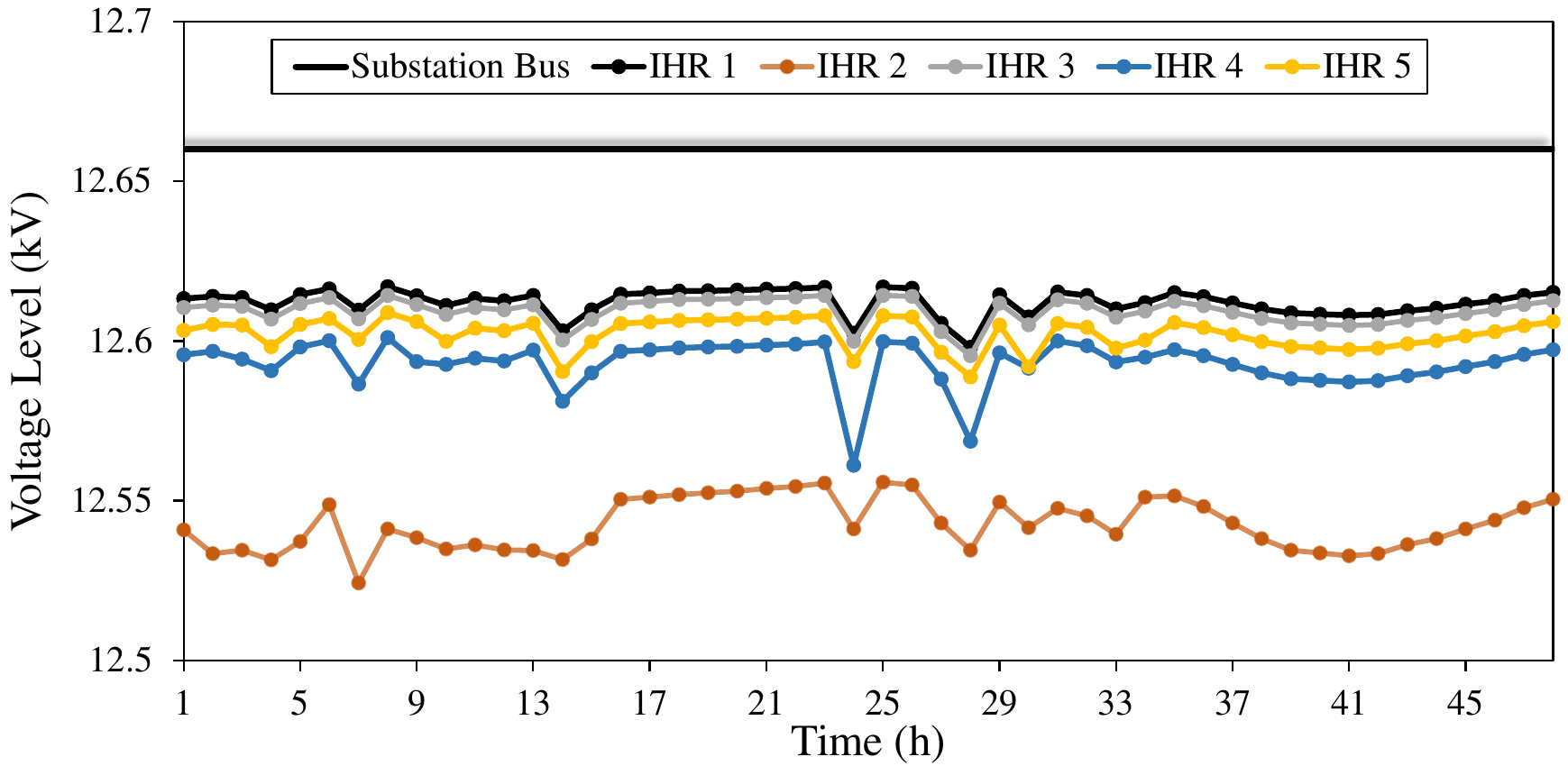}}
\caption{Voltage profiles of IHR nodes in the decentralized model, scenario $1$.} 
\label{fig:Voltage_IHR2}
\end{figure}

In the proposed hierarchical matching framework, the central agent solves a reduced-dimension OPF model with the IHRs as the nodes to deliver the IHR flow requirements while ensuring that the grid constraints are met.
The agent can also curtail the flow to each IHR to the extent that the grid constraints are not violated. Figure \ref{fig:Voltage_IHR2} shows the voltage profiles of IHRs in the decentralized model and scenario $1$.

\begin{figure}[ht!]
\centering{
\includegraphics[width=1\linewidth]{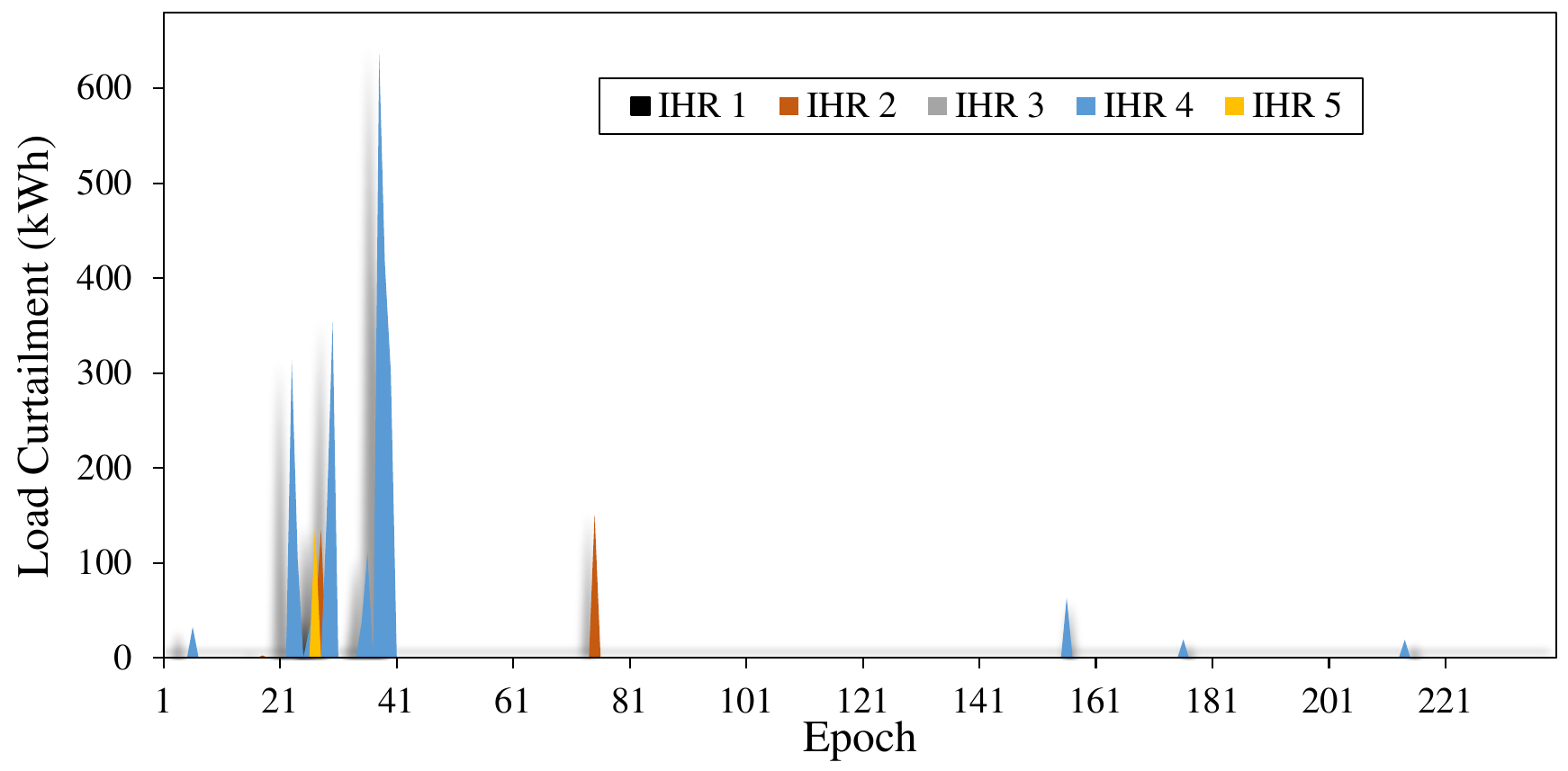}}
\caption{Matching curtailment of the central agent in the decentralized model, scenario $1$.} 
\label{fig:Curtail}
\end{figure}

In this epoch, the lower and upper voltage boundaries of IHRs are respectively $\underline{V}_{h}\!\!=\!\![12.37, 12.2, 12.05, 12.08, 12.22]$ and $\overline{V}_{h}\!\!=\!\![12.948, 13.11, 13.25, 13.22, 13.09]$.
As shown, the voltage level of all IHRs is within the safe lower-bound and upper-bound limits in all the time periods.
The extent to which the matching decisions are met in each IHR depends on the power flow in the grid operation, which can be curtailed by the central agent to ensure that the grid constraints are met. 
Figure \ref{fig:Curtail} shows the matching curtailment of different IHRs in the decentralized model and scenario 1. 
The results show that the initial IHR-level matching decisions are not curtailed in most of the epochs, though the matching decisions in some initial epochs are curtailed to ensure the safe operation of the power grid.     

\section{Conclusions} \label{Conclusions}
This paper proposes a learning-based hierarchical framework for dynamic matching in power distribution systems.
In the proposed framework, the power distribution system is divided into multiple IHRs, each consisting of flexible loads and RES.
The IHR agents employ DRL to output an efficient and scalable online matching policy to match the available RES and active customers as the day progresses, such that their quality of service constraints are not violated.
Once the IHR-level matching decisions are determined, a central agent uses the net active power flow, as well as the reactive power capacities of each IHR, to formulate a reduced-dimension OPF model to determine the final flows such that the flow requirements of the IHRs are met and the grid constraints are not violated.
The hierarchical approach offers a very effective way to combine the ability of DRL to learn state-dependent (or online) matching policies and that of optimization to ensure safe grid operation.
The proposed hierarchical framework was implemented on the IEEE 33-bus test distribution system and tested on multiple scenarios with different matching algorithms, including the proposed learning algorithm.
The results show that the hierarchical framework utilizes the flexible loads better, resulting in higher social welfare compared to the centralized approach that matches across the whole distribution system.

\bibliographystyle{IEEEtran}
\bibliography{ref}

\end{document}

%% file: macros.tex
%%%%%%%%%%%%%5
% new symbols
%\newcommand{\force}{\mbox{$\parallel \! \! \! -$}}
\newcommand{\force}{\mbox{$\Vdash$}}

% Hatted english letters
\newcommand{\ghat}{\mbox{$\bm \hat g$}}

\newcommand{\bzero}{\mbox{${\bm 0}$}}

% abbreviations for bold english letters (vectors)
\newcommand{\va}{\mbox{${\mathbf a}$}}
\newcommand{\vah}{\mbox{${\mathbf \hat a}$}}
\newcommand{\vat}{\mbox{${\mathbf \tilde a}$}}
\newcommand{\vb}{\mbox{${\mathbf b}$}}
\newcommand{\vd}{\mbox{${\mathbf d}$}}
\newcommand{\rh}{\mbox{${\hat r}$}}
\newcommand{\Itl}{\mbox{${\tilde I}$}}

\newcommand{\vxt}{\mbox{${\mathbf \tilde x}$}}
\newcommand{\vh}{\mbox{${\mathbf h}$}}
\newcommand{\vhh}{\mbox{${\mathbf \hat h}$}}
\newcommand{\ve}{\mbox{${\mathbf e}$}}
\newcommand{\vg}{\mbox{${\mathbf g}$}}
\newcommand{\vgh}{\mbox{${\mathbf \hat g}$}}
\newcommand{\vp}{\mbox{${\mathbf p}$}}
\newcommand{\vph}{\mbox{${\mathbf \hat p}$}}
\newcommand{\vq}{\mbox{${\mathbf q}$}}
\newcommand{\vt}{\mbox{${\mathbf t}$}}
\newcommand{\vw}{\mbox{${\mathbf w}$}}
\newcommand{\vwh}{\mbox{${\mathbf \hat w}$}}
\newcommand{\wh}{\mbox{${\hat w}$}}
\newcommand{\vwt}{\mbox{${\mathbf \tilde w}$}}
\newcommand{\wt}{\mbox{${\tilde w}$}}
\newcommand{\vs}{\mbox{${\mathbf s}$}}
\newcommand{\vsh}{\mbox{${\mathbf \hat s}$}}
\newcommand{\vst}{\mbox{${\mathbf \tilde s}$}}
\newcommand{\vr}{\mbox{${\mathbf r}$}}
\newcommand{\vx}{\mbox{${\mathbf x}$}}
\newcommand{\vv}{\mbox{${\mathbf v}$}}
\newcommand{\vu}{\mbox{${\mathbf u}$}}
\newcommand{\vy}{\mbox{${\mathbf y}$}}
\newcommand{\vz}{\mbox{${\mathbf z}$}}
\newcommand{\vn}{\mbox{${\mathbf n}$}}
\newcommand{\vnt}{\mbox{${\mathbf \tilde n}$}}
\newcommand{\vzero}{\mbox{${\mathbf 0}$}}
\newcommand{\vone}{\mbox{${\mathbf 1}$}}

% abbreviations for matrices
\newcommand{\mA}{\mbox{{$\mathbf A$}}}
\newcommand{\mAh}{\mbox{${\mathbf \hat A}$}}
\newcommand{\mAt}{\mbox{$\mathbf \tilde A$}}
\newcommand{\mB}{\mbox{${\mathbf B}$}}
\newcommand{\mBh}{\mbox{${\mathbf \hat B}$}}
\newcommand{\mC}{\mbox{{$\mathbf C$}}}
\newcommand{\mCh}{\mbox{${\mathbf \hat C}$}}
\newcommand{\mD}{\mbox{{$\mathbf D$}}}
\newcommand{\mDt}{\mbox{$\mathbf \tilde D$}}
\newcommand{\mE}{\mbox{{$\mathbf E$}}}
\newcommand{\mG}{\mbox{{$\mathbf G$}}}
\newcommand{\mF}{\mbox{{$\mathbf F$}}}
\newcommand{\mH}{\mbox{{$\mathbf H$}}}
\newcommand{\mHb}{\mbox{${\mathbf \bar H}$}}
\newcommand{\mI}{\mbox{{$\mathbf I$}}}
\newcommand{\mIh}{\mbox{${\mathbf \hat I}$}}
\newcommand{\mN}{\mbox{{$\mathbf N$}}}
\newcommand{\mM}{\mbox{{$\mathbf M$}}}
\newcommand{\mMh}{\mbox{{$\mathbf \hat M$}}}
\newcommand{\mP}{\mbox{${\mathbf P}$}}
\newcommand{\mQ}{\mbox{${\mathbf Q}$}}
\newcommand{\mR}{\mbox{${\mathbf R}$}}
\newcommand{\mRh}{\mbox{${\mathbf {\hat {R}}}$}}
\newcommand{\mRt}{\mbox{${\mathbf \tilde R}$}}
\newcommand{\mS}{\mbox{${\mathbf S}$}}
\newcommand{\mSb}{\mbox{${\mathbf \bar S}$}}
\newcommand{\mSh}{\mbox{${\mathbf \hat S}$}}
\newcommand{\mSt}{\mbox{${\mathbf \tilde S}$}}
\newcommand{\mT}{\mbox{${\mathbf T}$}}
\newcommand{\mU}{\mbox{${\mathbf U}$}}
\newcommand{\mUh}{\mbox{${\mathbf \hat U}$}}
\newcommand{\mV}{\mbox{${\mathbf V}$}}
\newcommand{\mVh}{\mbox{${\mathbf \hat V}$}}
\newcommand{\mW}{\mbox{${\mathbf W}$}}
\newcommand{\mWh}{\mbox{${\mathbf \hat W}$}}
\newcommand{\mWt}{\mbox{${\mathbf \tilde W}$}}
\newcommand{\mX}{\mbox{${\mathbf X}$}}
\newcommand{\mY}{\mbox{${\mathbf Y}$}}
\newcommand{\mZ}{\mbox{${\mathbf Z}$}}

%abbreviations for greek letters

\newcommand{\ga}{\alpha}
\newcommand{\gb}{\beta}
\newcommand{\grg}{\gamma}
\newcommand{\gd}{\delta}
\newcommand{\gre}{\varepsilon}
\newcommand{\gep}{\epsilon}
\newcommand{\gz}{\zeta}
\newcommand{\gzh}{\mbox{$ \hat \zeta$}}
\newcommand{\gh}{\eta}
\newcommand{\gth}{\theta}
\newcommand{\gi}{iota}
\newcommand{\gk}{\kappa}
\newcommand{\gl}{\lambda}
\newcommand{\gm}{\mu}
\newcommand{\gn}{\nu}
\newcommand{\gx}{\xi}
\newcommand{\gp}{\pi}
\newcommand{\gph}{\phi}
\newcommand{\gr}{\rho}
\newcommand{\gs}{\sigma}
\newcommand{\gsh}{\hat \sigma}
\newcommand{\gt}{\tau}
\newcommand{\gu}{\upsilon}
\newcommand{\gf}{\varphi}
\newcommand{\gc}{\chi}
\newcommand{\go}{\omega}

%Uppercase Greek

\newcommand{\gG}{\Gamma}
\newcommand{\gD}{\Delta}
\newcommand{\gTh}{\Theta}
\newcommand{\gL}{\Lambda}
\newcommand{\gX}{\Xi}
\newcommand{\gP}{\Pi}
\newcommand{\gS}{\Sigma}
\newcommand{\gU}{\Upsilon}
\newcommand{\gF}{\Phi}
\newcommand{\gO}{\Omega}

%Uppercase bold Greek

\def\bm#1{\mbox{\boldmath $#1$}}
\newcommand{\vga}{\mbox{$\bm \alpha$}}
\newcommand{\vgb}{\mbox{$\bm \beta$}}
\newcommand{\vgd}{\mbox{$\bm \delta$}}
\newcommand{\vge}{\mbox{$\bm \epsilon$}}
\newcommand{\vgl}{\mbox{$\bm \lambda$}}
\newcommand{\vgm}{\mbox{$\bm \mu$}}
\newcommand{\vgr}{\mbox{$\bm \rho$}}
\newcommand{\vgn}{\mbox{$\bm \nu$}}
\newcommand{\vgp}{\mbox{$\bm \pi$}}
\newcommand{\vgrh}{\mbox{$\bm \hat \rho$}}
\newcommand{\vgrt}{\mbox{$\bm {\tilde \rho}$}}

\newcommand{\vgt}{\mbox{$\bm \gt$}}
\newcommand{\vgth}{\mbox{$\bm {\hat \tau}$}}
\newcommand{\vgtt}{\mbox{$\bm {\tilde \tau}$}}
\newcommand{\vpsi}{\mbox{$\bm \psi$}}
\newcommand{\vphi}{\mbox{$\bm \phi$}}
\newcommand{\vxi}{\mbox{$\bm \xi$}}
\newcommand{\vth}{\mbox{$\bm \theta$}}
\newcommand{\vthh}{\mbox{$\bm {\hat \theta}$}}

\newcommand{\mgG}{\mbox{$\bm \Gamma$}}
\newcommand{\mgGh}{\mbox{$\hat {\bm \Gamma}$}}
\newcommand{\mgD}{\mbox{$\bm \Delta$}}
\newcommand{\mgU}{\mbox{$\bm \Upsilon$}}
\newcommand{\mgL}{\mbox{$\bm \Lambda$}}
\newcommand{\mPsi}{\mbox{$\bm \Psi$}}
\newcommand{\mgX}{\mbox{$\bm \Xi$}}
\newcommand{\mgS}{\mbox{$\bm \Sigma$}}

%Abbreviatians for blackboard bold
\newcommand{\oA}{{\open A}}
\newcommand{\oC}{{\open C}}
\newcommand{\oF}{{\open F}}
\newcommand{\oN}{{\open N}}
\newcommand{\oP}{{\open P}}
\newcommand{\oQ}{{\open Q}}
\newcommand{\oR}{{\open R}}
\newcommand{\oZ}{{\open Z}}

%Abbreviatians for caligraphic letters

\newcommand{\Nu}{{\cal V}}
\newcommand{\cA}{{\cal A}}
\newcommand{\cB}{{\cal B}}
\newcommand{\cC}{{\cal C}}
\newcommand{\cD}{{\cal D}}
\newcommand{\cF}{{\cal F}}
\newcommand{\cH}{{\cal H}}
\newcommand{\cK}{{\cal K}}
\newcommand{\cI}{{\cal I}}
\newcommand{\cL}{{\cal L}}
\newcommand{\cM}{{\cal M}}
\newcommand{\cN}{{\cal N}}
\newcommand{\cO}{{\cal O}}
\newcommand{\cP}{{\cal P}}
\newcommand{\cR}{{\cal R}}
\newcommand{\cS}{{\cal S}}
\newcommand{\cU}{{\cal U}}
\newcommand{\cV}{{\cal V}}
\newcommand{\cT}{{\cal T}}
\newcommand{\cX}{{\cal X}}

\newcommand{\rH}{^{*}}
\newcommand{\rT}{^{ \raisebox{1.2pt}{$\rm \scriptstyle T$}}}
\newcommand{\rF}{_{ \raisebox{-1pt}{$\rm \scriptstyle F$}}}
\newcommand{\rE}{{\rm E}}

% mathematical functions and constants
\newcommand{\dom}{\hbox{dom}}
\newcommand{\rng}{\hbox{rng}}
\newcommand{\Span}{\hbox{span}}
\newcommand{\Ker}{\hbox{Ker}}
\newcommand{\On}{\hbox{On}}
\newcommand{\otp}{\hbox{otp}}
\newcommand{\ZFC}{\hbox{ZFC}}
\def\Re{\ensuremath{\hbox{Re}}}
\def\Im{\ensuremath{\hbox{Im}}}
\newcommand{\SNR}{\ensuremath{\hbox{SNR}}}
\newcommand{\CRB}{\ensuremath{\hbox{CRB}}}
\newcommand{\diag}{\ensuremath{\hbox{diag}}}
\newcommand{\trace}{\ensuremath{\hbox{tr}}}

% more complex mathematical entities
\newcommand{\dlot}{\mbox{$\delta^1_3$}}
\newcommand{\Dlot}{\mbox{$\Delta^1_3$}}
\newcommand{\Dlof}{\mbox{$\Delta^1_4$}}
\newcommand{\dlof}{\mbox{$\delta^1_4$}}
\newcommand{\bP}{\mbox{$\mathbf{P}$}}
\newcommand{\Pot}{\mbox{$\Pi^1_2$}}
\newcommand{\Sot}{\mbox{$\Sigma^1_2$}}
\newcommand{\gDot}{\mbox{$\gD^1_2$}}

\newcommand{\Potr}{\mbox{$\Pi^1_3$}}
\newcommand{\Sotr}{\mbox{$\Sigma^1_3$}}
\newcommand{\gDotr}{\mbox{$\gD^1_3$}}

\newcommand{\Pofr}{\mbox{$\Pi^1_4$}}
\newcommand{\Sofr}{\mbox{$\Sigma^1_4$}}
\newcommand{\Dofr}{\mbox{$\gD^1_4$}}

\newcommand{\Sa}{\mbox{$S_{\ga}$}}
\newcommand{\Qk}{\mbox{$Q_{\gk}$}}
\newcommand{\Ca}{\mbox{$C_{\ga}$}}

\newcommand{\gkp}{\mbox{$\gk^+$}}
\newcommand{\aron}{ Aronszajn }

\newcommand{\sqkp}{\mbox{$\Box_{\gk}$}}
\newcommand{\dkp}{\mbox{$\Diamond_{\gk^{+}}$}}
\newcommand{\sqsqnce}
{\mbox{\\ $\ < \Ca \mid \ga < \gkp \ \ \wedge \ \ \lim \ga >$ \ \ }}
\newcommand{\dsqnce}{\mbox{$<S_{\ga} \mid \ga < \gkp >$}}

%%%%%%%%%%%%%%% Environnements %%%%%%%%%%%%%%%%%%
%\newcounter{theorem}
%\newtheorem{theorem}{Theorem}
%%\newtheorem{lemma}[theorem]{Lemma}
%\newtheorem{claim}[theorem]{Claim}
%\newcounter{definition}
%\newtheorem{definition}{Definition}

%%%%%%%%%%%%%%%%%%% Preambule %%%%%%%%%%%%%%%%%%%%

% mathematical environments
\newcommand{\beq}{\begin{equation}}
\newcommand{\eeq}{\end{equation}}
\newcommand{\bea}{\begin{array}}
\newcommand{\ena}{\end{array}}
\newcommand{\bds}{\begin {itemize}}
\newcommand{\eds}{\end {itemize}}
\newcommand{\bdf}{\begin{definition}}
\newcommand{\edf}{\end{definition}}
\newcommand{\blm}{\begin{lemma}}
\newcommand{\elm}{\end{lemma}}
\newcommand{\bthm}{\begin{theorem}}
\newcommand{\ethm}{\end{theorem}}
\newcommand{\bprp}{\begin{prop}}
\newcommand{\eprp}{\end{prop}}
\newcommand{\bcl}{\begin{claim}}
\newcommand{\ecl}{\end{claim}}
\newcommand{\bcr}{\begin{coro}}
\newcommand{\ecr}{\end{coro}}
\newcommand{\bquest}{\begin{question}}
\newcommand{\equest}{\end{question}}

%Abbreviatians for other symbols
\newcommand{\rarrow}{{\rightarrow}}
\newcommand{\Rarrow}{{\Rightarrow}}
\newcommand{\larrow}{{\larrow}}
\newcommand{\Larrow}{{\Leftarrow}}
\newcommand{\restrict}{{\upharpoonright}}
\newcommand{\nin}{{\not \in}}

%\newcommand{\ha}{\mbox{$\hat{\b{a}}$}}
%\newcommand{\tba}{\mbox{$\tilde{\b{a}}$}}
%\newcommand{\yh}{\mbox{${\hat y}$}}
%\newcommand{\vth}{\mbox{${\boldsymbol {\theta}}$}}
%\newcommand{\ups}{\mbox{${\Upsilon}$}}
%\newcommand{\vthh}{\mbox{${\boldsymbol {\hat \theta}}$}}
%\newcommand{\vtht}{\mbox{${\boldsymbol {\tilde \theta}}$}}
%\newcommand{\veps}{\mbox{${\underline \epsilon}$}}
%\newcommand{\vrho}{\mbox{${\boldsymbol {\rho}}$}}
%\newcommand{\vrhoh}{\mbox{${\boldsymbol {\hat \rho}}$}}
%\newcommand{\vrhot}{\mbox{${\boldsymbol {\tilde \rho}}$}}

%\newcommand{\vy}{\mbox{${\boldsymbol y}$}}
%\newcommand{\vyh}{\mbox{${\boldsymbol \check y}$}}
%\newcommand{\vyht}{\mbox{${\boldsymbol \hat y}$}}
%\newcommand{\vnh}{\mbox{${\boldsymbol \check n}$}}
%\newcommand{\omn}{\mbox{${\om_{_N}}$}}
%\newcommand{\pikon}{\mbox{${\frac{2\pi k}{N}}$}}

% Abrreviations
\newcommand{\ie}{\hbox{i.e.}}
\newcommand{\eg}{\hbox{e.g.}}